\documentclass[%
 reprint,
 amsmath,amssymb,
 aps,
prb,
]{revtex4-2}

\usepackage{physics}%for ket notation

\usepackage{graphicx}% Include figure files
\usepackage{dcolumn}% Align table columns on decimal point
\usepackage{bm}% bold math
\usepackage[colorlinks=true, linkcolor=black, citecolor=black, urlcolor=black]{hyperref}% add hypertext capabilities
\usepackage{float}
\usepackage{natbib}
\usepackage{xcolor}
\begin{document}

%\preprint{APS/123-QED}

\title{Optimal encoding of two dissipative interacting qubits}% Force line breaks with \\

\author{G. Di Bello$^{1}$}\author{G. De Filippis$^{2,3*}$}\author{A. Hamma$^{1,3*}$}\author{C. A. Perroni$^{2,3*}$} 
\affiliation{$^{1}$Dip. di Fisica E. Pancini - Università di Napoli Federico II - I-80126 Napoli, Italy}
\affiliation{$^{2}$SPIN-CNR and Dip. di Fisica E. Pancini - Università di Napoli Federico II - I-80126 Napoli, Italy
}
\affiliation{$^{3}$INFN, Sezione di Napoli - Complesso Universitario di Monte S. Angelo - I-80126 Napoli, Italy}

\begin{abstract}
We investigate a system of two coupled qubits interacting with an Ohmic bath as a physical model for the implementation of one logical qubit. In this model, the interaction with the other qubit represents unitary noise while the Ohmic bath is responsible for finite temperature. In the presence of a one-dimensional decoherence-free subspace (DFS), we show that, while this is not sufficient to protect a qubit from decoherence, it can be exploited to encode one logical qubit with greater performance than the physical one. We show different possible strategies for the optimal encoding of a logical qubit through a numerical analysis based on matrix product states. This method reproduces faithfully the results of perturbative calculations, but it can be extended to cases of crucial interest for physical implementations, e.g., in the case of strong coupling with the bath. As a result, a logical qubit encoded in the subspace which is the direct sum of the antiferromagnetic states in Bell basis, the DFS and the one in the triplet, is the optimally robust one, as it takes advantage of both the anchoring to the DFS and the protection from the antiferromagnetic interaction.

$^*$These authors contributed equally to this work, and their names are listed in alphabetical order.
\end{abstract}

\maketitle

%\tableofcontents
\section{Introduction}\label{sec:intro}
Optimal control of a qubit is of fundamental importance for the implementation of quantum technologies, the construction of reliable quantum devices, and quantum information processing \cite{kurizki2022thermodynamics,koch2022quantum,poggiali2018optimal}.

In the past two decades, there has been a significant increase of interest in the encoding of information in quantum computing, both from experimental and theoretical perspectives \cite{grace2006encoding, hodges2007experimental}. This surge of interest can be attributed to the ongoing need for improved techniques for optimal control in quantum systems. Researchers have developed and widely employed active and passive methods to protect quantum systems from decoherence and environmental effects. Active approaches involve the use of quantum error correction codes (QECCs) \cite{shor1995scheme}, while passive approaches are focused on noiseless quantum codes such as DFS or topological codes \cite{kitaev2003fault}. 

Extensive studies have focused on DFS implementation, including investigations of quantum registers \cite{zanardi1997noiseless} and the pairing of qubits with ancillas \cite{duan1997preserving}. Experimental demonstrations conducted during this time have showcased the ability of quantum computers to execute algorithms such as Shor's factorization \cite{mohseni2003experimental} and Grover's search \cite{ollerenshaw2003magnetic} by encoding information in DFS. Moreover, in \cite{fortunato2002implementation} universal control was demonstrated in a DFS-encoded qubit in a system of two nuclear spins using liquid state nuclear magnetic resonance techniques. 

%It's worth noting that they used a two-dimensional DFS subspace, since that the interaction they set only introduces bath dephasing effects.

However, several limitations have been identified regarding the use of DFS in real experimental implementations \cite{cappellaro2007control}. In practice, they depend on the fact that DFS are based on fine tuned symmetries that are hardly realistic. More realistically, though, the noise model may break most symmetries and shrink them to a point where they are useless.

%One notable limitation is the challenge of explicitly determining the form of the interaction between the system and the bath, raising concerns about accurate problem modeling. Additionally, the manipulation of quantum information within a DFS can be challenging due to the potential leakage of information caused by the presence of the bath, particularly when implementing operators that remain within the DFS during observation time .

In recent years, attempts have been made to generalize the idea of DFS and combine it with other methods, such as dynamical decoupling, to achieve better performance in quantum computation with spin chains \cite{qin2015protected}. Furthermore, researchers have explored the application of DFS in various systems, ranging from solid-state qubits implementations to wave-guide quantum electrodynamics \cite{lidar2014review, friesen2017decoherence, kockum2018decoherence}.

In this work, we study a model where a single qubit is interacting unitarily with another qubit, and both qubits are placed in a Ohmic bath. As a result, there is no two-dimensional DFS and therefore no protected qubit. However, a one-dimensional DFS survives. One may think of making a qubit with two such DFSs, but then one should take into account the coupling between these two degrees of freedom. Similarly, one can make one qubit by taking the DFS and another one-dimensional subspace in the space of two qubits. Both cases model the presence of additional unitary noise. In this work, we focus on the latter, more fundamental case, which includes the fact that once one couples a DFS to another subspace that is subject to dissipation, then the whole qubit is spoiled. 

And yet, one is tempted to think that anchoring part of the qubit to the DFS might increase its protection, as some of the decaying rates are cancelled, as it was shown in the perturbative case in \cite{campagnano2010entanglement}. In this work, we thoroughly study the optimal encoding of a logical qubit by exploiting the DFS, and using the MPS ansatz of tensor networks \cite{paeckel2019time,fishman2022itensor,haegeman2016unifying,haegeman2011time} numerical technique that allows us to go beyond the weak-coupling limit and consider system-bath interactions \cite{moroder2023stable,de2023signatures,flannigan2022many,strathearn2018efficient} that are more realistic and amenable to experimentation. 

We find that all the logical encodings that exploit the DFS perform better than the physical qubits. Moreover, we find that the optimal encoding results in the qubit made of the DFS and the other antiferromagnetic state in the triplet. In this way, both symmetry of the interaction with the bath and of the spin-spin interaction conspire to achieve the best performance.

We begin by summarizing the main results of this work in Sect.~\ref{sec:keyfind}, where we examine the logical subspace of the encoded qubit for three different strategies and compare it with the physical one. We demonstrate that the antiferromagnetic strategy is the optimal choice. In Sect.~\ref{sec:model}, we present the dissipative model comprising two interacting qubits in an Ohmic bath, along with the spectrum of the closed system Hamiltonian. Additionally, we analyze the effects of the bath on the free evolution of the open system for the physical qubits through the fidelity in Subsect.~\ref{subsec:batheff}. Next, in Sect.~\ref{sec:optimalenc}, we introduce three encoding strategies (Subsubsects.~\ref{subsubsec:sleaf}, \ref{subsubsec:mlesymm}, and \ref{subsubsec:mlenonsymm}) that encode information from the two physical qubits into one logical qubit. We proceed to optimize these strategies and compare their performance in Subsect.~\ref{subsec:comparison}, using fidelity and leakage (defined in the main text) as measures of the effect of the bath on the chosen encoding strategy. Then, we analyze what happens in the new logical subspace to the purity of the best strategies (Subsect.~\ref{subsec:purity}). Finally, in the concluding section, Sect.~\ref{sec:conclusion}, we summarize and further discuss our findings. In Appendix~\ref{app:lindblad}, we compare the Lindblad master equation solutions and the MPS numerical simulations. Furthermore, in Appendix~\ref{app:fidaf} we observe that also the fidelity in the logical subspace of the best encoded qubit remains very high during the evolution. Finally, in Appendix~\ref{app:bathcoup} we examine the impact of increasing system-bath coupling on the best encoding strategy we have identified.

\section{Overview of the key findings}\label{sec:keyfind}
\begin{figure}[H]
    \begin{center}
        \includegraphics[scale=0.26]{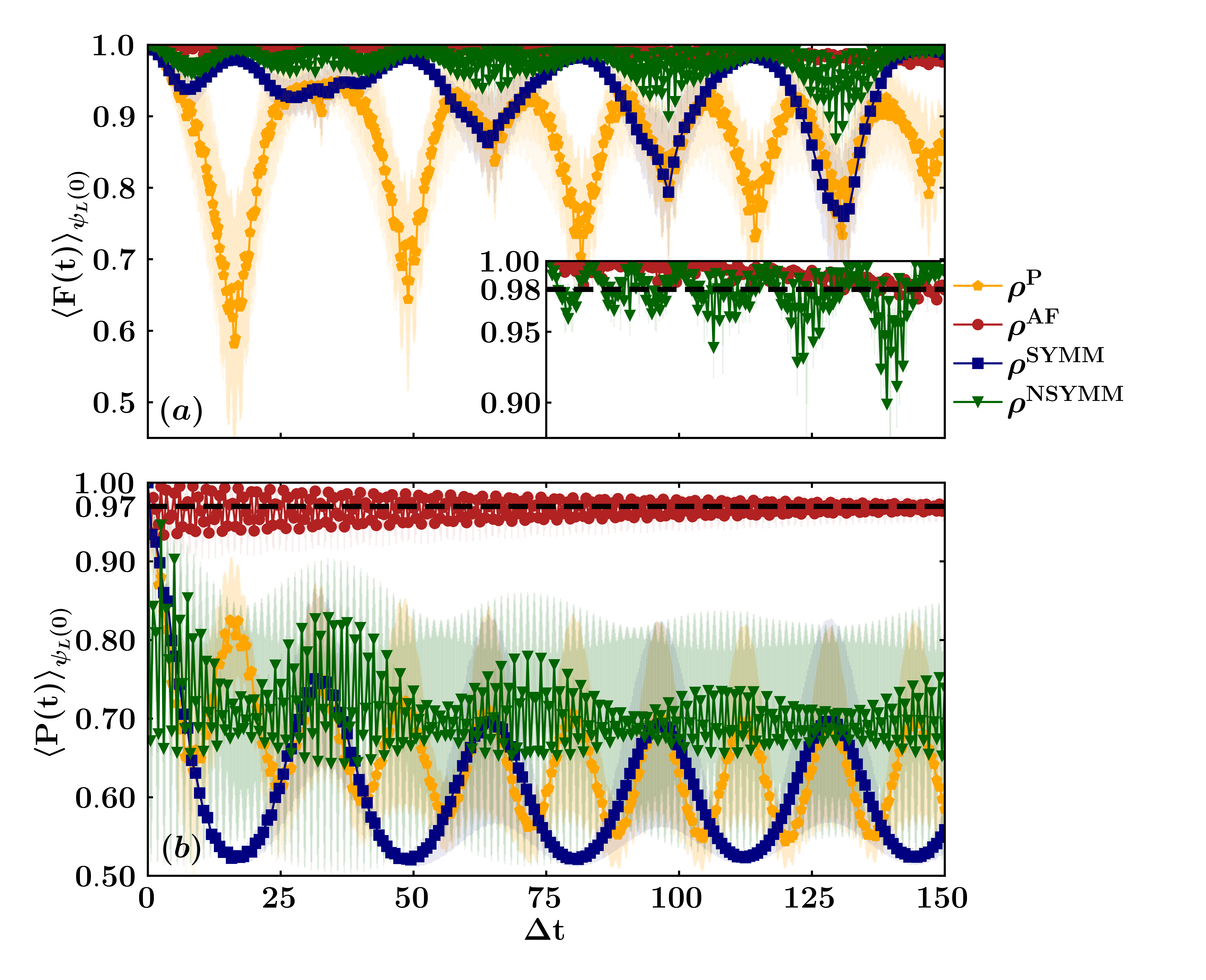}  
        \caption{\label{fig:plotiniz}Fidelity $F(t)$ of the free evolution with the open system evolution ($a$) and purity $P(t)$ ($b$) for encoded qubits AF (red circle), SYMM (blue square), NSYMM (green triangle) (more details on the definition in the main text) and one physical qubit (orange pentagon) as functions of dimensionless time $\Delta t$ for $\nu=-5\Delta$, $\alpha=0.01$, and the initial state $\ket{\psi_L(0)}=\cos{\theta}\ket{\uparrow_L}+e^{i\phi}\sin{\theta}\ket{\downarrow_L}$ where $\theta$ and $\phi$ sample all the logical qubit Hilbert space. We average the fidelity and the purity over $18$ realizations of these angles. The dots in the plots correspond to the average value of the fidelity and the purity, while the shaded regions around them show the range of values covered by the standard deviation of the fidelity and the purity.} 
    \end{center}
\end{figure}

In this work, we present strategies (AF, SYMM, NSYMM, see Sect.~\ref{sec:optimalenc}) for encoding information in a qubit subjected to both unitary and non-unitary noises. We demonstrate that by optimizing with respect to the fidelity of free evolution compared to evolution in the presence of the environment, these strategies outperform simply encoding information in a physical qubit. This is evident in Figure~\ref{fig:plotiniz}, where we operate within the new logical subspaces chosen to implement these encoding strategies and average over the pure states located on the surface of the new logical Bloch sphere. The figure illustrates fidelity (Figure~\ref{fig:plotiniz}.a), which measures the effect of the bath on the encoded qubit's dynamics, and purity (Figure~\ref{fig:plotiniz}.b), which remains near 1, ensuring us that the new qubit remains in a quantum state.\\
Furthermore, from the plots, we can discern that the best strategy is the AF (Antiferromagnetic) strategy, with more details provided in Sect. \ref{subsubsec:sleaf}. In fact, the fidelity consistently oscillates around a value of 0.98 (see the inset in Figure~\ref{fig:plotiniz}.a) throughout our simulation time, while the purity exhibits a similar trend for extended time intervals around an asymptotic value of 0.97. This exceptional performance of the AF strategy stems from our utilization of the two antiferromagnetic states in the Bell basis for the up and down states, harnessing the power of the DFS and the antiferromagnetic interaction ($\nu=-5\Delta$). Consequently, we establish a quasi-one-dimensional DFS that exhibits remarkable resilience to bath effects over extended durations and in the presence of a significantly strong coupling to the environment.

\section{Dissipative two interacting qubits model}\label{sec:model} 
\begin{figure}[H]
    \begin{center}
        \includegraphics[scale=0.105]{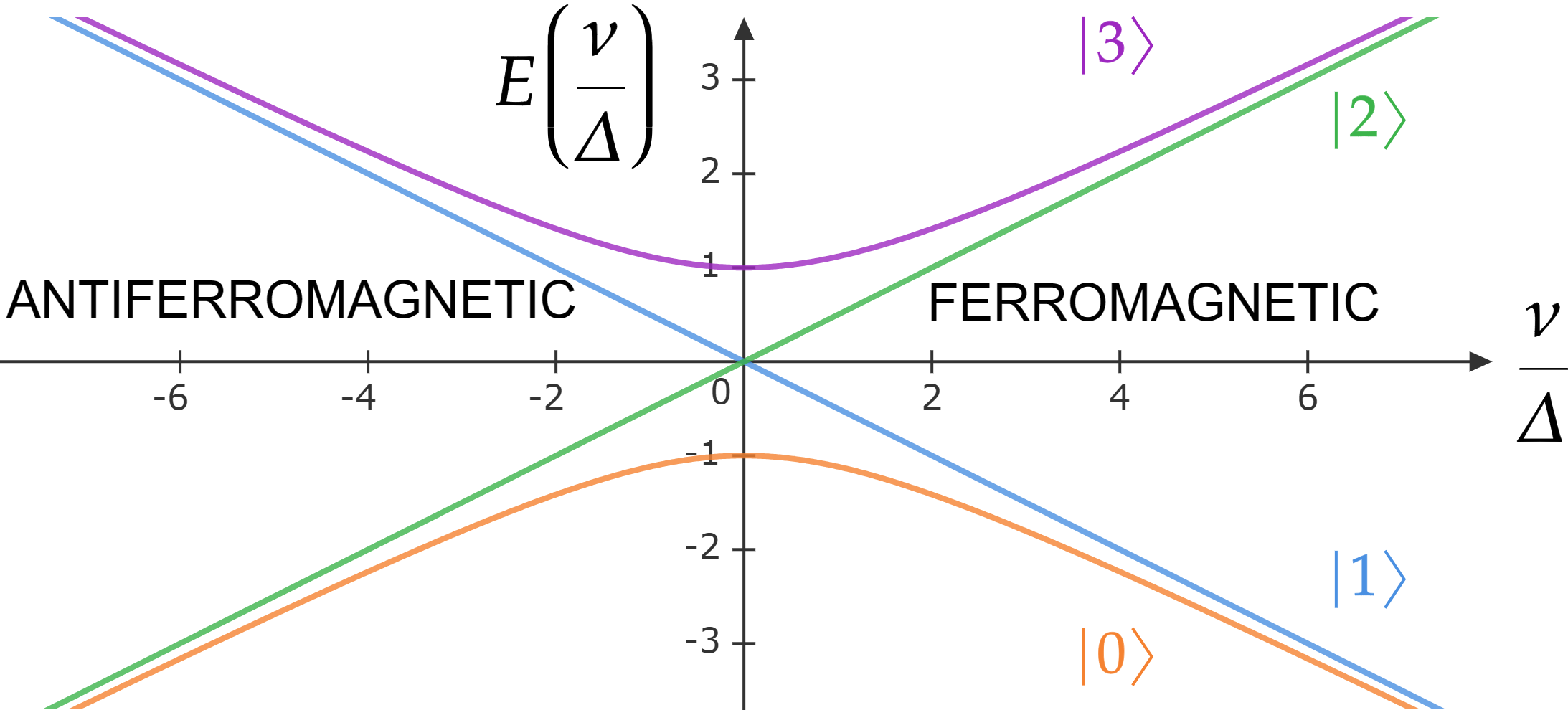}
     
        \caption{\label{fig:spettro}Energy spectrum (in units of $\Delta$) of the closed system consisting of two interacting qubits, as described by the Hamiltonian $H_{qub}$, as a function of the ratio $\nu/\Delta$.} 
    \end{center}
\end{figure}

We are considering a system of two interacting qubits immersed in an Ohmic common bath, which is composed of a set of $N$ harmonic oscillators. The model we are considering assumes a zero temperature thermal bath and we set $\hbar=1$. The Hamiltonian that describes the system is given by:
\begin{equation}
    \label{Hgen}
    H=H_{qub} + H_{bath} + H_{qub-bath}.    
\end{equation}
Here, the qubits energy $H_{qub}$ is 
\begin{equation}
\label{eq:Hqub}
    H_{qub}= -\frac{\Delta}{2}(\sigma_x^1+\sigma_x^2)-\frac{\nu}{2}\sigma_z^1\sigma_z^2,    
\end{equation}
with $\Delta$ the frequency of the two qubits, labelled with the superscripts $^{1}$ or $^{2}$, $\nu$ the strength of the interaction between them and $\sigma_i^j$, with $i=x,y,z$ and $j=1,2$, the Pauli matrices. The bath Hamiltonian $H_{bath}$ is 
\begin{equation}
    H_{bath}= \sum_{i=1}^{N} \omega_{i} a^{\dagger}_i a_i,    
\end{equation}
with $\omega_i$ the frequencies of the bath modes and $a_i$ ($a^{\dagger}_i$) the annihilation (creation) operators for the $N$ harmonic oscillators. The qubits-bath interaction Hamiltonian $H_{qub-bath}$ is given by:
\begin{equation}
    H_{qub-bath}=(\sigma_z^1+\sigma_z^2)\sum_{i=1}^{N} \lambda_i (a_i+a^{\dagger}_i),   
\end{equation}
with $\lambda_i$ the couplings between each qubit and each bath mode. We set the coupling in this way because the long-wavelength limit applies to the case of two qubits \cite{fradkin2013field}. The coupling to the bath $\alpha$ is defined in such a way that the bath spectral density $J(\omega)$ can be written as follows:
\begin{equation}
\label{eq:specdens}
    J(\omega)=\sum_{i=1}^N \mid \lambda_i \mid^2 \delta(\omega-\omega_i)=\frac{\alpha}{2}\omega f\left(\frac{\omega}{\omega_c}\right),
\end{equation}
where $f\left(\frac{\omega}{\omega_c}\right)$ is a function depending on the cutoff frequency for the bath modes $\omega_c$, ruling the behavior of the spectral density at high frequencies. This function can be $f\left(\frac{\omega}{\omega_c}\right)=\Theta\left(\frac{\omega_c}{\omega}-1\right)$, where $\Theta(x)$ is the Heaviside step function or an exponential decay $f\left(\frac{\omega}{\omega_c}\right)=e^{-\frac{\omega}{\omega_c}}$. This cutoff frequency is typically chosen to be of the order of the largest energy scale in the system.\\
To gain a deeper understanding of how states affect the dynamic behavior of the system, we begin by diagonalizing the closed Hamiltonian $H_0=H_{qub}$. The resulting spectrum is shown in Figure~\ref{fig:spettro}:
\begin{equation}
\label{eq:eigen}
    \frac{E\left(\frac{\nu}{\Delta}\right)}{\Delta}=\left\{-\sqrt{1+\left(\frac{\nu}{2\Delta}\right)^2};\,-\frac{\nu}{2\Delta};\, \frac{\nu}{2\Delta};\,\sqrt{1+\left(\frac{\nu}{2\Delta}\right)^2}\right\}.
\end{equation}

The corresponding four eigenstates of the closed system can be written in the Bell basis \cite{nielsen2010quantum} $\left\{\ket{S}\equiv\ket{\Psi^-},\,\,\ket{T,AF}\equiv\ket{\Psi^+},\,\,\ket{T,F+}\equiv\ket{\Phi^+},\right.$
$\left.\ket{T,F-}\equiv\ket{\Phi^-}\right\}$, 
where $S$ stands for ``singlet", while $T$ for ``triplet", and the specifications $AF$ and $F$, respectively, for ``antiferromagnetic" and ``ferromagnetic".
Hence, the Hamiltonian eigenstates read:
\begin{align}
    \ket{0}=&a\left(\frac{\nu}{\Delta}\right)\ket{T,AF}-b\left(\frac{\nu}{\Delta}\right)\ket{T,F+}
    \\
    \ket{1}=&\ket{T,F-}\\
    \ket{2}=&\ket{S}\\
    \ket{3}=&a\left(\frac{\nu}{\Delta}\right)\ket{T,F+}+b\left(\frac{\nu}{\Delta}\right)\ket{T,AF},\label{eq:excited_eigenstate}
\end{align}
where $0,1,2,3$ go from the ground state ($0$) to the most excited one ($3$). The two coefficients $a$ and $b$ depend only on the parameters of the Hamiltonian and are defined as:

\begin{align}
\label{eq:afac}
    a\left(\frac{\nu}{\Delta}\right)&=\frac{2}{\sqrt{4+\left(\frac{\nu}{\Delta}+\sqrt{4+\left(\frac{\nu}{\Delta}\right)^2}\right)^2}}\\
    \label{eq:bfac}
    b\left(\frac{\nu}{\Delta}\right)&=-\frac{\sqrt{4+\left(\frac{\nu}{\Delta}\right)^2}+\frac{\nu}{\Delta}}{\sqrt{4+\left(\frac{\nu}{\Delta}+\sqrt{4+\left(\frac{\nu}{\Delta}\right)^2}\right)^2}}.
\end{align}

To thoroughly explore the entire Hilbert space and identify the optimal encoding strategy that remains independent of the initial state, we opt to sample from a uniform distribution. Specifically, we set the initial state as a linear combination of the singlet $\ket{S}$ and the triplet in the Bell basis $\left\{\ket{T,AF},\ket{T,F+},\ket{T,F-}\right\}$ as follows:
\begin{align}
    \ket{\psi(0)}&=d_S\ket{S}+d_{T,AF}\ket{T,AF}\nonumber\\
    &+d_{T,F+}\ket{T,F+}+d_{T,F-}\ket{T,F-}.
\end{align}
Here, the coefficients $d_{T,AF},d_{T,F+},$ and $d_{T,F-}$ are complex numbers, but we choose to make a discretization such that their squared magnitudes can take on one of the values $\{0.0,0.25,0.5,0.75,1.0\}$, and their phases can be one of the values $\left\{0,\frac{\pi}{2},\pi,\frac{3}{2}\pi\right\}$. This results in $M=(5\times4)^4=20^4=160,000$ potential initial states. To eliminate redundancy, we remove the inessential global phase, arbitrarily setting the first coefficient $d_S$ to be real. States where the magnitude is $0$ but the phase differs from $0$ are also discarded, as they lead to double-counting. Additionally, configurations violating the normalization condition, where the sum of the squared magnitudes of the four coefficients deviates from $1$, are omitted. Through a code designed to check these conditions, we reduce the count from $M=160,000$ to $M=332$ different realizations of the initial state, over which we compute the average fidelity over time. \\
In order to verify if our sampling of the entire Hilbert space is sufficiently faithful, we compute the quantity:
\begin{equation}
    \label{eq:sample}
    \left\vert\left\vert \frac{1}{M}\sum_{i=1}^M \rho_i(0) - \frac{\mathrm{I}}{4} \right\vert\right\vert_F,
\end{equation}
where $M$ is the number of realizations, in our case $332$, $\rho_i(0)$ is the initial state of the i-th realization, $\mathrm{I}$ is the identity matrix, and the used matrix norm is the Frobenius one. We expect this norm to be zero if the sampling is correct, and indeed it is zero within our numerical precision ($\leq 10^{-15}$).

\subsection{Bath effects on the qubits}\label{subsec:batheff}
\begin{figure}[H]
    \begin{center}
        \includegraphics[scale=0.26]{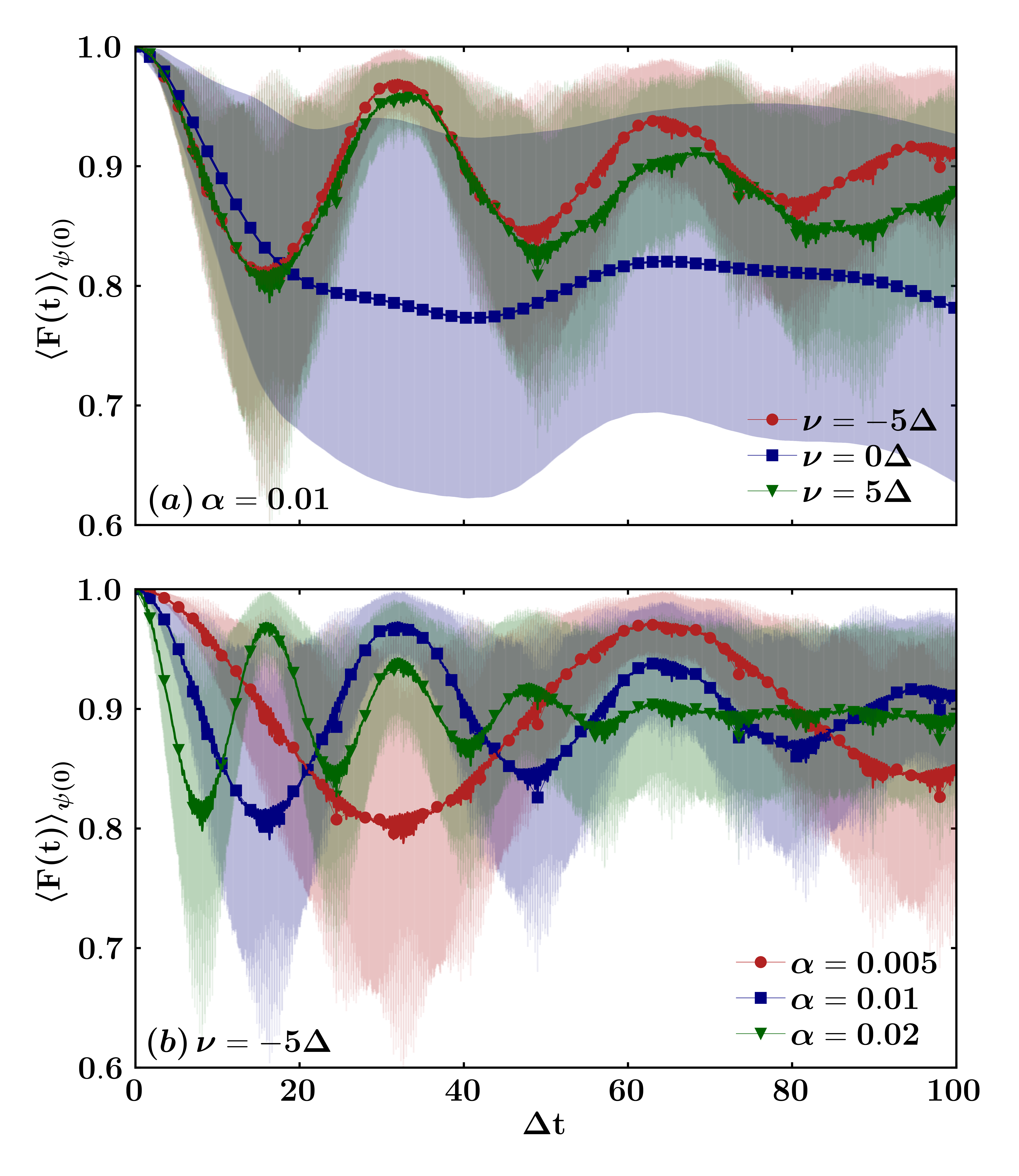} 
        \caption{\label{fig:fid_initialst1}Fidelity $F(t)$ of the free evolution of one physical qubit with the open system evolution as a function of dimensionless time $\Delta t$ for $\alpha=0.01$ and $\nu=-5\Delta$ (red circle), $\nu=0\Delta$ (blue square), $\nu=5\Delta$ (green triangle) in $(a)$ and for $\nu=-5\Delta$ and $\alpha=0.005$ (red circle), $\alpha=0.01$ (blue square), $\alpha=0.02$ (green triangle) in $(b)$. The initial state is $\ket{\psi(0)}=d_S\ket{S}+d_{T,AF}\ket{T,AF}+d_{T,F+}\ket{T,F+}+d_{T,F-}\ket{T,F-}$, where $d_S$, $d_{T,AF}$, $d_{T,F+}$ and $d_{T,F-}$ sample all the two-qubits Hilbert space. We average the fidelity over $332$ realizations of these coefficients. The dots in the plots correspond to the average value of the fidelity, while the shaded regions around them in light colors show the range of values covered by the standard deviation of the fidelity.} 
    \end{center}
\end{figure}

Our goal is to investigate the influence of the bath on the physical qubits and the encoded qubit. To achieve this, we compute the Uhlmann fidelity $F$ of the density matrix of the open system, $\rho^o(t)$, with respect to that of the closed system (without bath interaction) $\rho^c(t)$ \cite{nielsen2010quantum} by using MPS numerical simulations (Appendix~\ref{app:lindblad} and Appendix~\ref{app:MPS}):
\begin{equation}
    F\left[\rho^o(t),\rho^c(t)\right]=Tr\left\{\sqrt{\sqrt{\rho^o(t)}\rho^c(t)\sqrt{\rho^o(t)}}\right\}.
\end{equation}
We use $\Delta$ as the energy unit and set the value of the coupling to the bath, $\alpha$, in the range of $[0.005, 0.02]$ (Appendix~\ref{app:bathcoup}). This range allows us to analyze the typical coupling regime where the effect of the bath can be detrimental \cite{de2023signatures,di2023qubit}. We also analyze the system for different values of $\nu$, specifically $\nu = \{-5, 0, 5\}\Delta$.\\
When examining the fidelity of a single physical qubit over time, averaged over all possible initial states (see Figure~\ref{fig:fid_initialst1}), we observe a minimum value at short times of about $0.8$ for the mean value and $0.6$, considering the standard deviation, for all values of $\alpha$ and $\nu$. In particular, when varying the interaction strength $\nu$ (see Figure~\ref{fig:fid_initialst1}.a), we observe that the worst-case scenario occurs for $\nu=0\Delta$, both for the mean value of the oscillations and the standard deviation. This effect stems from the interaction with the bath. The qubits tend to reach their ground state earlier due to entanglement with the bath rather than with each other. Consequently, information flows towards the bath, causing the fidelity to decrease as the qubits significantly differ from a closed-system scenario. The behavior is similar for the other two values of $\nu$. However, in the ferromagnetic case, there is a decreasing mean value of the oscillations, while in the antiferromagnetic case, the mean value is almost constant and higher than the others. Figure~\ref{fig:fid_initialst1}.b shows that decreasing $\alpha$ results in the same fidelity behavior, but with lower oscillation frequencies, thereby shortening the period. Roughly speaking, a doubling of $\alpha$ leads to a halving of the period. By extrapolating the period's behavior concerning $\alpha$, we observe a quasi-hyperbolic relationship represented as $T=a/\alpha$, where fitting the three points to such a curve suggests $a\approx1/3$. This finding is consistent with the limit of $\alpha$ approaching zero, where the fidelity must remain constantly one, and therefore, the oscillation frequency is zero and the period tends towards infinity. Given this worst-case behavior for a single physical qubit, we need to determine the optimal way to encode information into a logical qubit.\\
For the remainder of the paper, we focus on the results obtained for $\alpha=0.01$ and $\nu=-5\Delta$, as they are the most interesting and promising. We point out that the value of the coupling is within the intermediate regime, for which we need non-perturbative methods to analyze the dynamics. This coupling strength value takes advantage of the antiferromagnetic interaction, which can help mitigate the effects of the bath (the exact elimination of the bosonic degrees of freedom induces an effective ferromagnetic time retarded interaction between the two spins). In Appendix~\ref{app:bathcoup}, we demonstrate how increasing the coupling to the bath affects the optimal encoding strategy.

\section{Optimal encodings in one logical qubit}\label{sec:optimalenc}
The general idea is to encode the logical up and down states in other states. In particular, we choose these states to be the Bell states or a linear combination of them. We define the logical up $\ket{\psi_{\uparrow_L}}$ and logical down $\ket{\psi_{\downarrow_L}}$ as follows:
\begin{align}
    \ket{\psi_{\uparrow_L}} &= \sum_{i=1}^{N_{\uparrow}} l_i \ket{\psi_i} \\
    \ket{\psi_{\downarrow_L}} &= \sum_{i=1}^{N_{\downarrow}} k_i \ket{\phi_i},
\end{align}
where $N_{\uparrow}$ and $N_{\downarrow}$ are the dimensions of the subspaces chosen for the encoding, and $\psi_i$ and $\phi_i$ are the states that span the subspaces, respectively\cite{grace2006encoding}. The normalization conditions are satisfied such that $\sum_{i=1}^{N_{\uparrow}}|l_i|^2=1$ and $\sum_{i=1}^{N_{\downarrow}}|k_i|^2=1$.\\
With this choice, the generic state of the encoded qubit is given by: $\ket{\psi^E} = \alpha \ket{\psi_{\uparrow_L}} + \beta \ket{\psi_{\downarrow_L}}$, where $\alpha$ and $\beta$ are complex coefficients satisfying the normalization condition $|\alpha|^2+|\beta|^2=1$. To represent the encoded qubit, we need to find the new Pauli operators $\sigma_i^{E}$, where $i=x,y,z$. These operators are given by:
\begin{align}
    \sigma_x^{E}&=\ket{\psi_{\uparrow_L}}\bra{\psi_{\downarrow_L}}+\ket{\psi_{\downarrow_L}}\bra{\psi_{\uparrow_L}}\\
    \sigma_y^{E}&=-i\ket{\psi_{\uparrow_L}}\bra{\psi_{\downarrow_L}}+i\ket{\psi_{\downarrow_L}}\bra{\psi_{\uparrow_L}}\\
    \sigma_z^{E}&=\ket{\psi_{\uparrow_L}}\bra{\psi_{\uparrow_L}}-\ket{\psi_{\downarrow_L}}\bra{\psi_{\downarrow_L}}.
\end{align} 
These operators allow us to define the encoded qubit state:
\begin{equation}
    \rho^{E}(t)=\frac{1}{2}\begin{pmatrix}
        1+\langle\sigma_z^{E}(t)\rangle & \langle\sigma_x^{E}(t)\rangle -i\langle\sigma_y^{E}(t)\rangle\\
        \langle\sigma_x^{E}(t)\rangle +i\langle\sigma_y^{E}(t)\rangle & 1-\langle\sigma_z^{E}(t)\rangle
    \end{pmatrix}.
\end{equation}

\subsection{Encoding strategies and optimization}
Our approach involves utilizing the singlet and triplet subspaces to encode one logical qubit in the two physical qubits. In the following sections, we present our three proposed encoding strategies and their respective optimizations.

\subsubsection{Antiferromagnetic single-level encoding}\label{subsubsec:sleaf}
Due to the presence of an antiferromagnetic interaction ($\nu=-5\Delta$), here we utilize a single-level encoding (SLE) strategy to encode information in the two physical qubits. Specifically, we select the singlet state $\ket{S}$ as the logical up state, which is also an eigenstate of the closed Hamiltonian $\left(\ket{2}\right)$, and the antiferromagnetic state of the triplet in the Bell basis $\ket{T,AF}$ as the logical down state. This encoding strategy, which we refer to as ``antiferromagnetic" encoding, works by exploiting the properties of the singlet state and the antiferromagnetic interaction to create a sort of semi-DF-subspace encoding, since it almost coincides with the ground state manifold (see Figure~\ref{fig:spettro}). In addition, upon examining the analytical expression of fidelity within the Lindblad approach, it is confirmed that the it is the most resilient one, remaining constantly 1 in the limit of $|\nu|>>\Delta$ (see Appendix D). The singlet state, indeed, is decoherence-free, while the antiferromagnetic state of the triplet in the Bell basis is the state that is most protected by the antiferromagnetic interaction. By encoding information in these two states, we can mitigate errors and noise caused by the bath, which makes our system more robust and reliable. 

\subsubsection{Symmetric multilevel encoding} \label{subsubsec:mlesymm}
\begin{figure}[H]
    \begin{center}
        \includegraphics[scale=0.075]{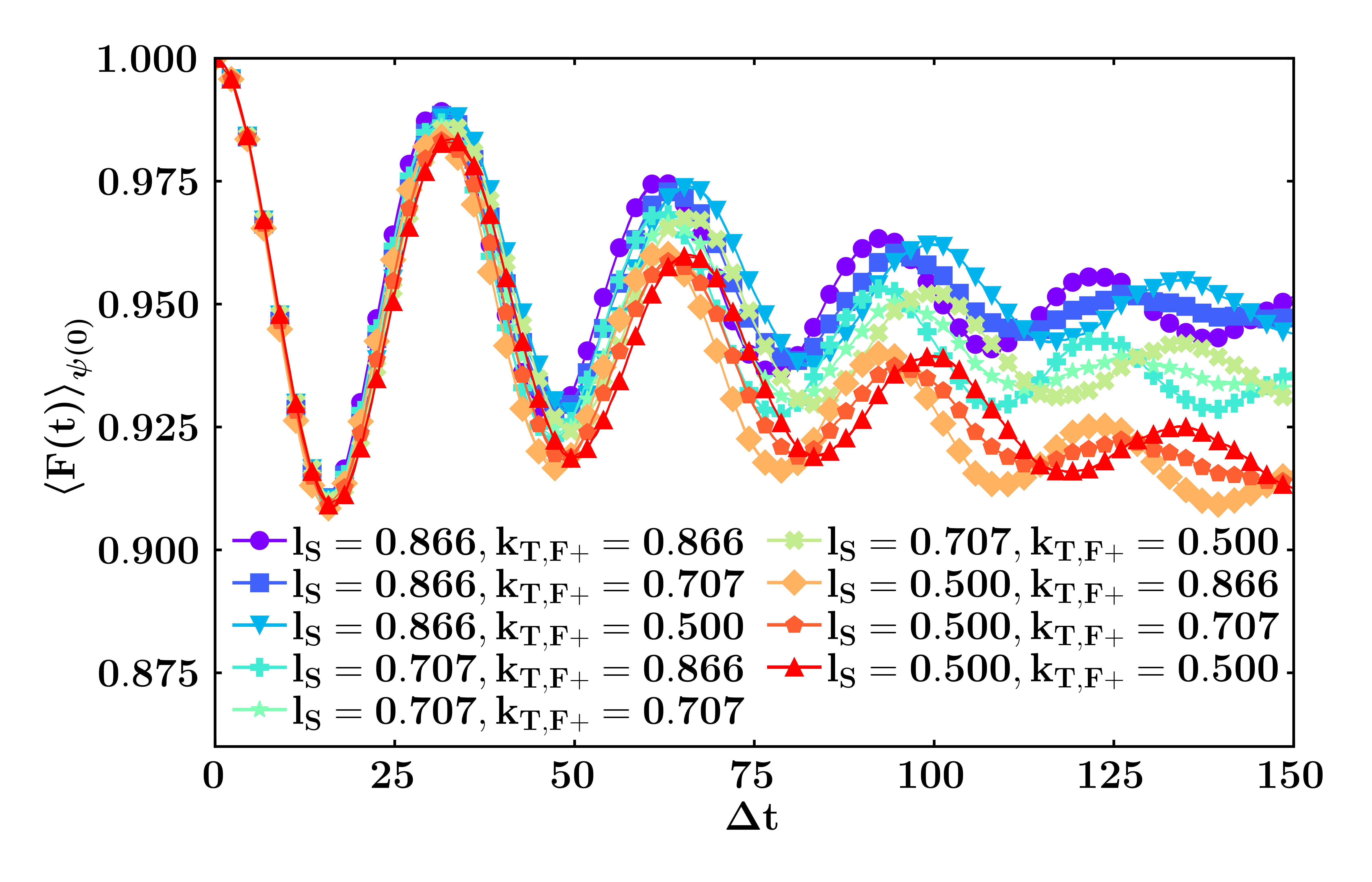} 
        \caption{\label{fig:fid_encsimm}Fidelity $F(t)$ of the free evolution of the encoded qubits OPTSYMM with the open system evolution for different values of $l_S^{SM}$ and $k_{T,F+}^{SM}\in \left\{0.25,0.50,0.75\right\}$ (rainbow colours from violet to red) as a function of dimensionless time $\Delta t$ for $\nu=-5\Delta$, $\alpha=0.01$, and the initial state $\ket{\psi(0)}=d_S\ket{S}+d_{T,AF}\ket{T,AF}+d_{T,F+}\ket{T,F+}+d_{T,F-}\ket{T,F-}$, where $d_S$, $d_{T,AF}$, $d_{T,F+}$ and $d_{T,F-}$ sample all the two-qubits Hilbert space. We average the fidelity over $332$ realizations of these coefficients.} 
    \end{center}
\end{figure}

\begin{figure}[H]
    \begin{center}
        \includegraphics[scale=0.27]{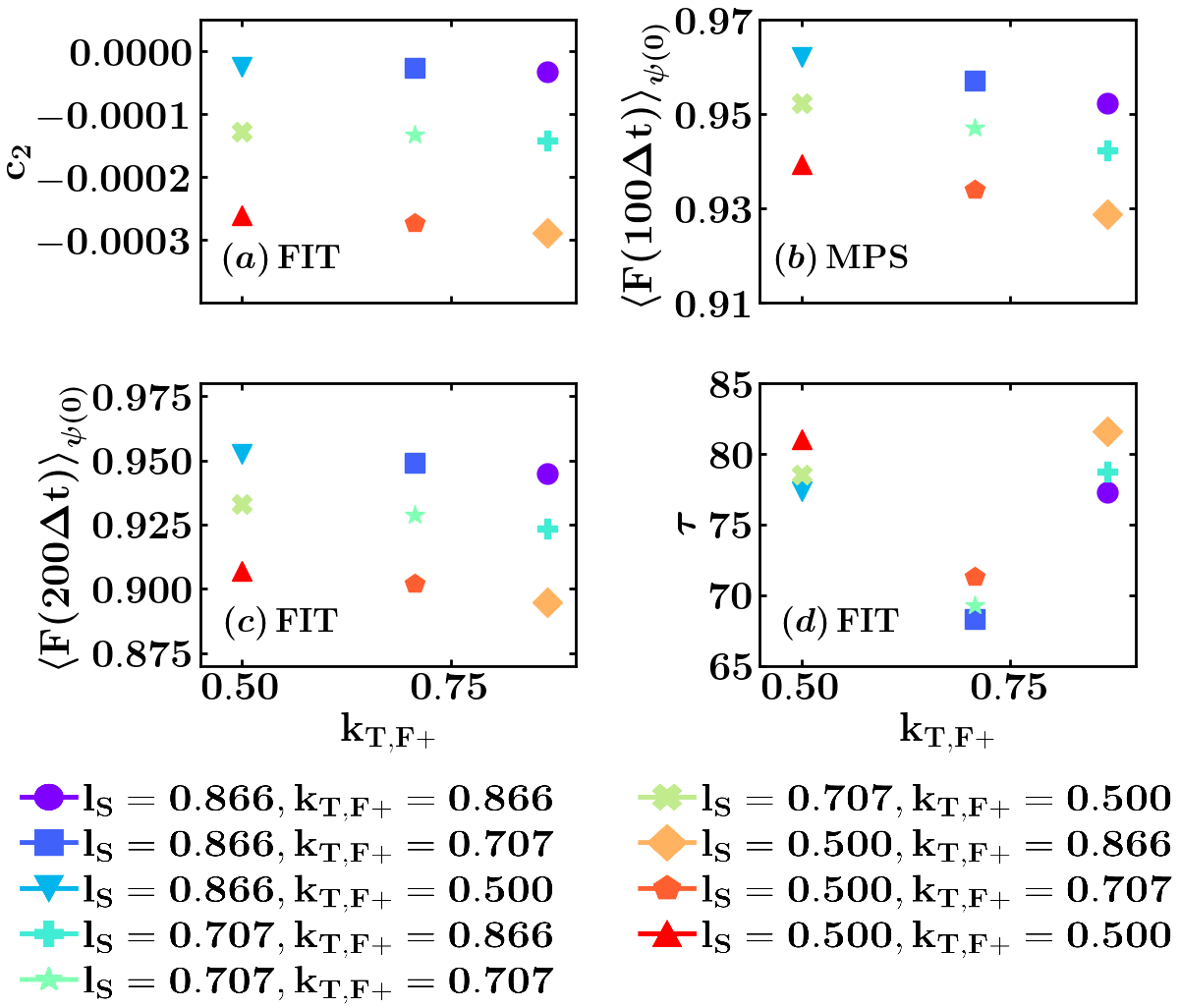}  
        \caption{\label{fig:fid_fit_simm}Fidelity$F(t)$ of the free evolution of the encoded qubits OPTSYMM with the open system evolution for different values of $l_S^{SM}$ and $k_{T,F+}^{SM}\in \left\{0.25,0.50,0.75\right\}$ (rainbow colours from violet to red) as a function of dimensionless time $\Delta t$ for $\nu=-5\Delta$, $\alpha=0.01$, and the initial state $\ket{\psi(0)}=d_S\ket{S}+d_{T,AF}\ket{T,AF}+d_{T,F+}\ket{T,F+}+d_{T,F-}\ket{T,F-}$, where $d_S$, $d_{T,AF}$, $d_{T,F+}$ and $d_{T,F-}$ sample all the two-qubits Hilbert space. We average the fidelity over $332$ realizations of these coefficients. In the plot $(a)$ there is the parameter $c_2$ responsible of the linear behavior, while in $(d)$ we observe the decay rate $\tau=1/c_4$. Plots $(b)$ and $(c)$ show the fidelity from MPS data at $\Delta\,t=100$ and from fitting data at $\Delta\,t=200$ respectively.} 
    \end{center}
\end{figure}

Here we consider as a possible implementation of the multilevel encoding (MLE) strategy the ``symmetric" case in which use an equal number of states to span the two logical subspaces. \\
In particular, we select two antiferromagnetic states from the Bell basis, $\ket{S}$ and $\ket{T,AF}$, to span the logical up subspace. Similarly, we choose two ferromagnetic states, $\ket{\psi_{F,+}}$ and $\ket{\psi_{F,-}}$, to span the logical down subspace. The logical up and down can be expressed as:
\begin{align}
    \ket{\psi_{\uparrow_L}^{SM}} &= l_S^{SM}\ket{S} + l_{T,AF}^{SM}\ket{T,AF}\\
    \ket{\psi_{\downarrow_L}^{SM}} &= k_{T,F+}^{SM}\ket{T,F+} + k_{T,F-}^{SM}\ket{T,F-}
\end{align}
where $l_S^{SM}$ and $l_{T,AF}^{SM}$ are coefficients representing the contributions of the logical up states, and $k_{T,F+}^{SM}$ and $k_{T,F-}^{SM}$ are coefficients representing the contributions of the logical down states. It's important to note that these states are normalized, so $|l_S^{SM}|^2 +|l_{T,AF}^{SM}|^2=|k_{T,F+}^{SM}|^2+|k_{T,F-}^{SM}|^2=1$. We also vary these coefficients to find the best encoding strategy.\\
We first check that the phases of the coefficients have only a slight effect on the oscillations of the standard deviation of the fidelity. Therefore, we set the coefficients to be real. We vary the squared modulus of $l_S^{SM}$ and $k_{T,F+}^{SM}$ in the interval $\left\{0.25,0.50,0.75\right\}$, and the other coefficients are determined by normalization, resulting in an ``optimal symmetric" encoding (OPTSYMM). We then observe the effect of this encoding on the fidelity of the qubit over time. Figure~\ref{fig:fid_encsimm} shows that increasing $l_{s}^{SM}$ (or decreasing $l_{T,AF}^{SM}$) leads to a higher asymptotic value of the fidelity. When $l_S^{SM}$ and $l_{T,AF}^{SM}$ are fixed, varying $k_{T,F+}^{SM}$ and $k_{T,F-}^{SM}$ only affects the frequency of the oscillations, with a slight improvement in fidelity for the lower value of $k_{T,F+}^{SM}$, which corresponds to the main contribution to the most excited eigenstate of the closed Hamiltonian. \\
To optimize the encoding strategy, we fit the curve in our range of time with four parameters $c_i$, where $i=1, \dots, 4$, using the following function:
\begin{equation}
\label{eq:fit}
    F_{FIT}\left[\rho^o(t),\rho^c(t)\right]= c_1 + c_2 t + (1-c_1)\cos(c_3 t) e^{-c_4 t}.
\end{equation}
This function takes into account the linear behavior of the average ($c_2$) and the damped oscillations, where $c_3$ determines the frequency of the oscillations and $c_4$ determines the rate of decay. The parameter $c_1$ ensures that the fitted curve starts at the value of $1$, which is defined as the absence of any bath effect. \\
To determine the best encoding parameters and coefficients, we analyze how the fitting parameters depend on $l_S^{SM}$ and $k_{T,F+}^{SM}$ and how they relate to the fidelity over time. To examine the asymptotic behavior of the fidelity, we take its value from the numerical data at time $\Delta t=100$ and extrapolate it from the fit at time $\Delta t=200$ (respectively Figure~\ref{fig:fid_fit_simm}.b and \ref{fig:fid_fit_simm}.c).\\
We can stop our analysis at $\Delta t=200$ because the large $\tau=1/c_4$ value from the fit, which quantifies the decay, is approximately $75$, as shown in Figure~\ref{fig:fid_fit_simm}.d. Therefore, we have already observed the complete behavior of the system over the time interval of $\Delta t=200$.\\
We observe that the higher the value of $l_S^{SM}$, the higher the fidelity of the encoding strategy becomes. If we further increase its contribution, the $c_2$ coefficient passes through zero and becomes slightly positive, although it remains on the order of $10^{-4}$. Therefore, we choose $l_S^{SM}=\sqrt{3}/2$ and $k_{T,F+}^{SM}=1/2$ as the best encoding in the symmetric case. 

\subsubsection{Nonsymmetric multilevel encoding }\label{subsubsec:mlenonsymm}
\begin{figure}[H]
    \begin{center}
        \includegraphics[scale=0.25]{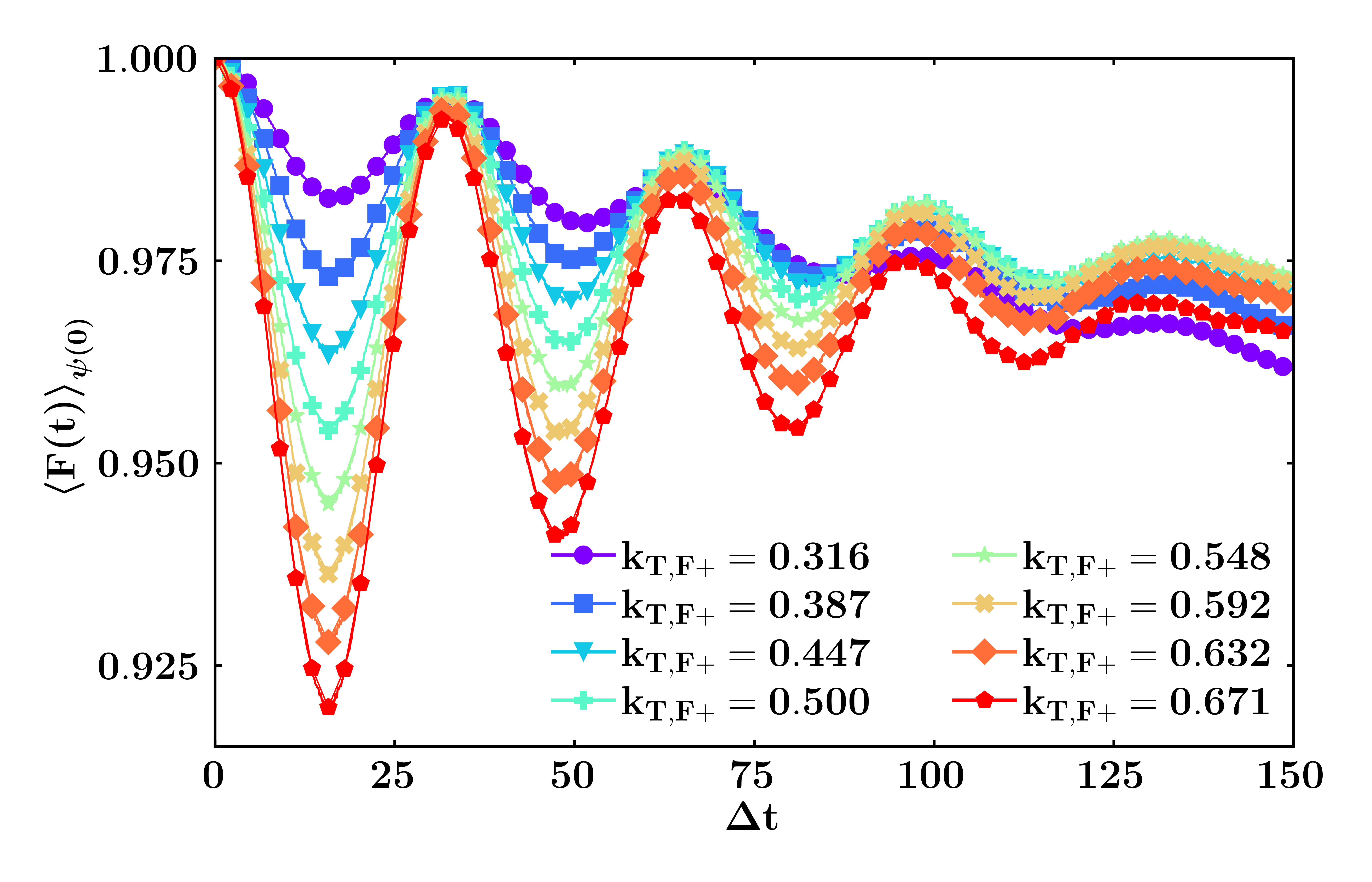}  
        \caption{\label{fig:fid_enc2a1}Fidelity $F(t)$ of the free evolution of the encoded qubits OPTNSYMM with the open system evolution for increasing values of $k_{T,F+}^{NSM}\in \left]0,1/\sqrt{2}\right[$ (rainbow colours from violet to red) as a function of dimensionless time $\Delta t$ for $\nu=-5\Delta$, $\alpha=0.01$, and the initial state $\ket{\psi(0)}=d_S\ket{S}+d_{T,AF}\ket{T,AF}+d_{T,F+}\ket{T,F+}+d_{T,F-}\ket{T,F-}$, where $d_S$, $d_{T,AF}$, $d_{T,F+}$ and $d_{T,F-}$ sample all the two-qubits Hilbert space. We average the fidelity over $332$ realizations of these coefficients.} 
    \end{center}
\end{figure}

\begin{figure}[b]
    \begin{center}
        \includegraphics[scale=0.38]{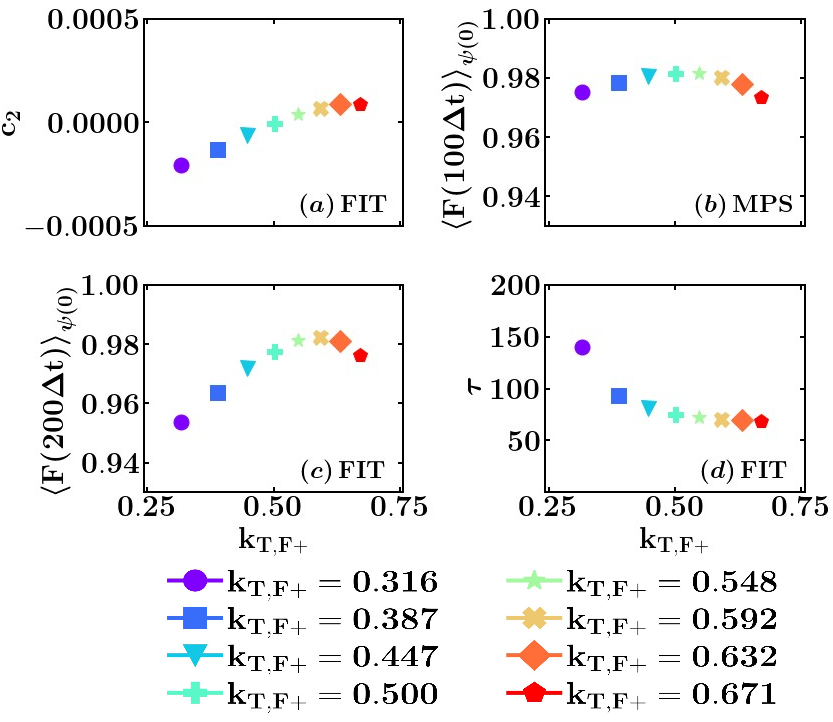}
        \caption{\label{fig:fid_fit}Fidelity $F(t)$ of the free evolution of the encoded qubits OPTNSYMM with the open system evolution for increasing values of $k_{T,F+}^{NSM}\in \left]0,1/\sqrt{2}\right[$ (rainbow colours from violet to red) as a function of dimensionless time $\Delta t$ for $\nu=-5\Delta$, $\alpha=0.01$, and the initial state $\ket{\psi(0)}=d_S\ket{S}+d_{T,AF}\ket{T,AF}+d_{T,F+}\ket{T,F+}+d_{T,F-}\ket{T,F-}$, where $d_S$, $d_{T,AF}$, $d_{T,F+}$ and $d_{T,F-}$ sample all the two-qubits Hilbert space. We average the fidelity over $332$ realizations of these coefficients. In the plot $(a)$ there is the parameter $c_2$ responsible of the linear behavior, while in $(d)$ we observe the decay rate $\tau=1/c_4$. Plots $(b)$ and $(c)$ show the fidelity from MPS data at $\Delta\,t=100$ and from fitting data at $\Delta\,t=200$ respectively.} 
    \end{center}
\end{figure}

The second implementation of the MLE strategy that we introduce here is the ``nonsymmetric" encoding, in which we use a different number of states to span each subspace. In particular, we encode the logical up in the singlet state $\ket{\psi_{\uparrow_L}^{NSM}}=\ket{S}$ ($l_S^{NSM}=1$), while varying the logical down in the triplet. Specifically, the logical down can be written as an arbitrary combination of the three states of the triplet in the Bell basis as follows:
\begin{equation}
    \ket{\psi_ {\downarrow_L}^{NSM}}=k_{T,AF}^{NSM}\ket{T,AF}+k_{T,F+}^{NSM}\ket{T,F+}+k_{T,F-}^{NSM}\ket{T,F-},
\end{equation}
and it is normalized: $|k_{T,AF}^{NSM}|^2+|k_{T,F+}^{NSM}|^2+|k_{T,F-}^{NSM}|^2=1$. 
We optimize the coefficients of $\ket{\psi_{\downarrow_L}^{NSM}}$ to improve the fidelity of the encoded qubit. As before, we set the coefficients to be real. We also find that the coefficient $k_{T,AF}^{NSM}$ has the most significant contribution to the fidelity, while the other two coefficients ($k_{T,F+}^{NSM}$ and $k_{T,F-}^{NSM}$) related to the higher system eigenenergies are less important. Hence, we set them to be equal and the other coefficient is determined by normalization, making an ``optimal nonsymmetric" encoding (OPTNSYMM). This allows us to vary a single parameter $k_{T,F+}^{NSM}\in \left]0,1/\sqrt{2}\right[$ and observe the effect on the fidelity over time of the encoded qubit. In this way, we explore the cases in between the AF strategy (i.e., $k_{T,AF}^{NSM}=1$ and $k_{T,F+}^{NSM}=k_{T,F-}^{NSM}=0$), and the case in which $k_{T,F+}^{NSM}=k_{T,F-}^{NSM}=1/\sqrt{2}$, completely canceling the contribution of the antiferromagnetic triplet state. Figure~\ref{fig:fid_enc2a1} shows that increasing $k_{T,F+}^{NSM}$ (decreasing $k_{T,AF}^{NSM}$) leads to a lower minimum value of the fidelity at short times, but a higher asymptotic value. Using the same fitting function (Eq.~\ref{eq:fit}), we find the best compromise between these two behaviors.
As previously done for the symmetric encoding, we observe how the fitting parameters depend on $k_{T,F+}^{NSM}$. We concentrate on the range of coefficients that offer the optimal trade-off between the minimum short-term value and the highest asymptotic value, specifically centered around $c_2$ passing through zero, indicating a constant average value (see Figure~\ref{fig:fid_fit}.a). Again, we take the fidelity from the numerical data at time $\Delta t=100$ and extrapolate it from the fit at time $\Delta t=200$ (respectively Figure~\ref{fig:fid_fit}.b and \ref{fig:fid_fit}.c) to analyze its asymptotic behavior. \\
The best encoding with the corresponding fit is obtained for $k_{T,F+}^{NSM}\approx 1/\sqrt{3}$. 

\subsection{Comparison of optimal encodings: fidelity and leakage}\label{subsec:comparison}
After optimizing the two MLE strategies, we aim to compare them with AF and determine the best approach.\\
We begin by computing the fidelity of the three encoded qubits and the physical one. For the encoded ones we fix the coefficients of $\ket{\psi_{\uparrow_L}}$ and $\ket{\psi_{\downarrow_L}}$ such that they can be:

\begin{itemize}
    \item ``antiferromagnetic" encoding (AF), SLE in Bell basis $\left(\ket{\psi_{\uparrow_L}}=\ket{S}\right.$ and $\left.\ket{\psi_{\downarrow_L}}=\ket{T,AF}\right)$ (see Subsubsect.~\ref{subsubsec:sleaf});
    \item ``symmetric" encoding (SYMM), MLE in Bell basis $\left(\ket{\psi_{\uparrow_L}}=\frac{\sqrt{3}}{2}\ket{S}+\frac{1}{2}\ket{T,AF}\right.$ and $\left.\ket{\psi_{\downarrow_L}}=\frac{1}{2}\ket{T,F+}+\frac{\sqrt{3}}{2}\ket{T,F-}\right)$ (see Subsubsect.~\ref{subsubsec:mlesymm});
    \item ``nonsymmetric" encoding (NSYMM), MLE in Bell basis $\bigg(\ket{\psi_{\uparrow_L}}=\ket{S}$ and $\left.\ket{\psi_{\downarrow_L}}=\frac{\ket{T,AF}+\ket{T,F+}+\ket{T,F-}}{\sqrt{3}}\right)$ (see Subsubsect.~\ref{subsubsec:mlenonsymm}).
\end{itemize}

\begin{figure}[H]
    \begin{center}
        \includegraphics[scale=0.27]{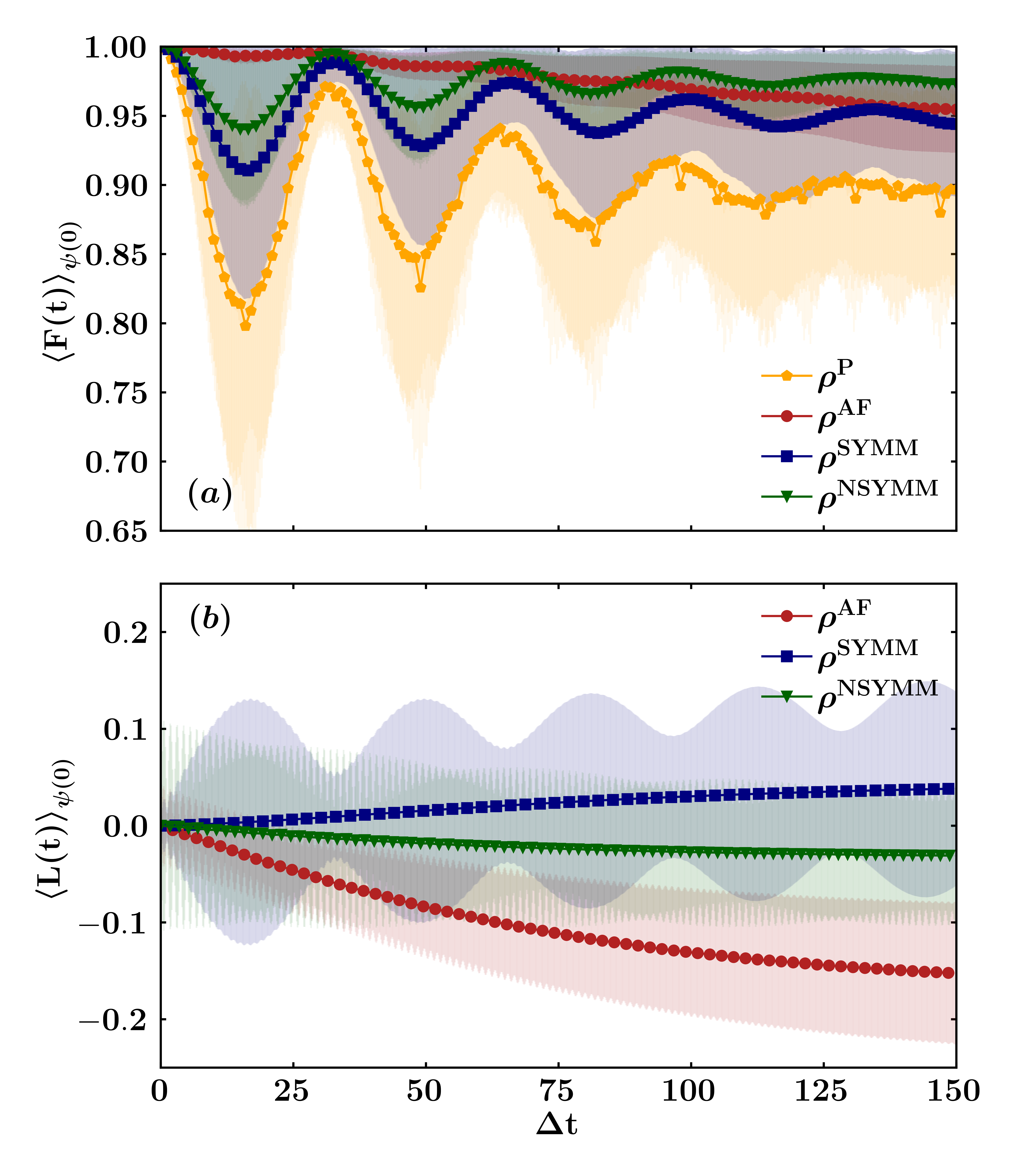}
        \caption{\label{fig:fid_leak_varienc}Fidelity $F(t)$ of the free evolution with the open system evolution ($a$) and leakage $L(t)$ ($b$) of encoded qubits AF (red circle), SYMM (blue square) and NSYMM (green triangle) (more details on the definition in the main text) and one physical qubit (orange pentagon) as functions of dimensionless time $\Delta t$ for $\nu=-5\Delta$, $\alpha=0.01$, and the initial state $\ket{\psi(0)}=d_S\ket{S}+d_{T,AF}\ket{T,AF}+d_{T,F+}\ket{T,F+}+d_{T,F-}\ket{T,F-}$, where $d_S$, $d_{T,AF}$, $d_{T,F+}$ and $d_{T,F-}$ sample all the two-qubits Hilbert space. We average the fidelity and the leakage over $332$ realizations of these coefficients. The dots in the plots correspond to the average value of the fidelity and the leakage, while the shaded regions around them show the range of values covered by the standard deviation of the fidelity and the leakage.} 
    \end{center}
\end{figure}

 We assess the fidelity of the three encoding strategies, and the results are presented in Figure~\ref{fig:fid_leak_varienc}.a. Our findings show that all encoding strategies outperform the physical qubit. Specifically, the AF strategy exhibits higher fidelity at short times, but its fidelity linearly decreases over time with small damped oscillations. On the other hand, allowing for a non-zero probability for other states in the two logical states initially worsens fidelity, but ultimately reaches a better asymptotic value. The two MLE strategies demonstrate a higher asymptotic value for fidelity. The symmetric strategy has a short-time minimum and an asymptotic value lower than the nonsymmetric case. This behavior can be explained by the system dynamics, which involves a timescale during which higher-order processes with the bath become dynamically allowed, and low-energy states can be excited, overcoming the energy gap. As the effect of the bath increases, higher-energy levels decay, leading to a stationary state that ultimately increases the fidelity. Therefore, including all triplet states in one of the encoded logical states allows for better accounting of the various processes arising during the dynamics.
Also, we introduce a quantity called ``leakage" that measures the effect of the bath (i.e., the environment) on the encoding strategy. We can define the leakage as follows:
\begin{equation}
\label{eq:leakage}
    L(t) = \frac{\left| Tr_L\left\{\Pi_Q\rho(0)\right\}\right|-\left|Tr_L\left\{\Pi_Q^U(t)\rho(0)\right\}\right|}{2\sqrt{2}}.
\end{equation}
In this definition, $\Pi_Q$ is the projector in the encoded space, which is spanned by the new logical up and down. $\Pi_Q$ is defined as the outer product of the two basis states $\ket{\psi_{\uparrow_L}}$ and $\ket{\psi_{\downarrow_L}}$. The superscript $U$ denotes the evolution under the complete Hamiltonian, and hence $\Pi_Q^U$ is the evolved projector. The trace is performed in the encoded logical space, which is indicated by the subscript $L$. The factor $2\sqrt{2}$ in the denominator is a normalization factor such that the leakage assumes only values in the range $[-1;1]$. This quantity is a measure of how dissipation makes the system deviate from or return to the logical subspace during the evolution.\\
Figure~\ref{fig:fid_leak_varienc}.b indicates that the NSYMM and the AF strategies are the best options, being negative, which means that the dynamics increases the populations in the logical subspace. However, the leakage for AF strategy reaches a lower asymptotic value. \\
Based on our analysis, we can conclude that the AF SLE encoding strategy in the Bell basis allows for better information extraction from two physical qubits to one logical qubit, while remaining robust to the detrimental effects of the bath over time (Appendix~\ref{app:bathcoup}). Additionally, we conducted this analysis for both sub-Ohmic and super-Ohmic spectral densities. Our findings consistently demonstrate that the AF strategy remains the optimal choice regardless of the bath's specific form.

\subsection{Encoded qubit: analysis of the purity}\label{subsec:purity}

\begin{figure}[H]
    \begin{center}
        \includegraphics[scale=0.26]{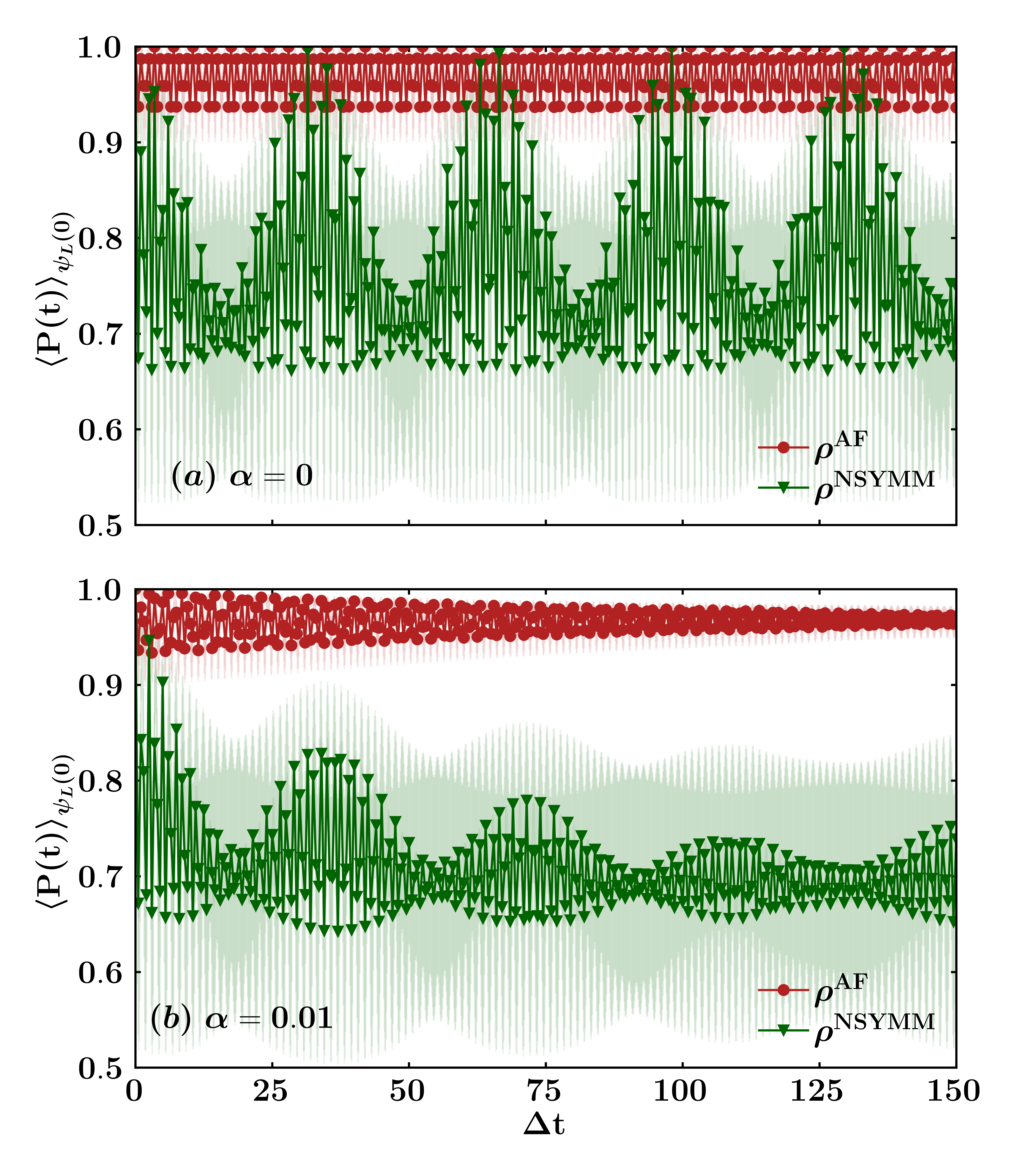}  
        \caption{\label{fig:pur_bestenc}Purity $P(t)$ of encoded qubits for $\alpha=0$ ($a$) and $\alpha=0.01$ ($b$) as a function of dimensionless time $\Delta t$ for $\nu=-5\Delta$, for the AF strategy (red circle), and NSYMM strategy (green triangle), and the initial state $\ket{\psi_L(0)}=\cos{\theta}\ket{\uparrow_L}+e^{i\phi}\sin{\theta}\ket{\downarrow_L}$ where $\theta$ and $\phi$ sample all the logical qubit Hilbert space. We average the purity over $18$ realizations of these angles. The dots in the plots correspond to the average value of the purity, while the shaded regions around them show the range of values covered by the standard deviation of the purity.} 
    \end{center}
\end{figure}

We are now prepared to examine the newly encoded logical subspace to determine whether the encoded qubit retains its quantum properties, as indicated by the purity of its state. To do this, we will perform an average over 18 realizations of the initial state, which is extracted from a uniform distribution within the new logical encoded space. This initial state is defined as a linear combination of the two new logical states, up and down, as follows:
\begin{equation}
    \ket{\psi_L(0)}=\cos{\theta}\ket{\uparrow_L}+e^{i\phi}\sin{\theta}\ket{\downarrow_L}.
\end{equation}
Similar to our approach for averaging initial states in the two-qubit case, we will verify that the sampling of the logical Hilbert space provides an adequate representation of the problem. Our analysis indicates that the norm in (\ref{eq:sample}) is indeed approximately $10^{-8}$.\\
Since in Sect.~\ref{sec:keyfind} in Figure~\ref{fig:plotiniz}, we describe all three encoding strategies compared with the physical one and find that the best two approaches are NSYMM and AF, as identified in the previous section, in what follows we will focus on these two. \\
In Figure~\ref{fig:pur_bestenc}.a, we depict the qubits in isolation from the bath. It is evident that the purity exhibits periodic oscillations, periodically returning to a value of 1, signifying the preservation of quantum characteristics in our qubit. Notably, the AF strategy exhibits smaller amplitude oscillations, closer to a purity value of 1.\\
Conversely, Figure~\ref{fig:pur_bestenc}.b presents the scenario where there is a non-zero coupling to the bath. In the case of NSYMM, all peaks are shorter, and the purity no longer reaches a value of 1 over time; instead, it converges towards a stationary value of approximately 0.7. In contrast, for AF, although the amplitude of oscillations is reduced, the purity consistently remains high, stabilizing at around 0.97.\\
Based on our analysis, we can confidently conclude that the AF encoding strategy is an excellent choice for encoding information that is highly resilient to the effects of the bath over time. 

\section{Discussion and conclusions}\label{sec:conclusion}
In the earlier section, we established that the most effective encoding strategy, following our analysis, is the antiferromagnetic SLE (Subsubsect.~\ref{subsubsec:sleaf}). This strategy involves the antiferromagnetic state of the triplet in the logical down state and the singlet state as the logical up state. This superior performance might be attributed to the use of the states closer to the lowest levels of the system thanks to the antiferromagnetic interaction between the qubits. It outperforms the nonsymmetric MLE strategy (Subsubsect.~\ref{subsubsec:mlenonsymm}) that is also a good choice, in particular in terms of fidelity. In this case we can observe the excitation of low-energy states and the decay of higher-energy levels, owing to higher-order processes, which lead to a stationary state that enhances fidelity. In other words, as suggested in \cite{grace2006encoding} mixing and decoherence in the encoding subspace do not
affect the coherence between the logical states and therefore do not reduce the computational
fidelity. Specifically, the nonsymmetric case outperforms the symmetric one (Subsubsect.~\ref{subsubsec:mlesymm}) because it do not couple the two subspaces (singlet and triplet) and also leverages the DFS (singlet).\\
In the previous sections, we focused on the strength of the antiferromagnetic interaction $\nu=-5\Delta$ and the system-bath coupling $\alpha=0.01$. However, we have conducted a comprehensive analysis. Regarding the physical qubits shown in Figure~\ref{fig:fid_initialst1}.a, the best performance is achieved with the antiferromagnetic interaction, which is true for all encoding strategies. In fact, the qubits, when influenced by the bath, tend to establish a ferromagnetic order. Therefore, an antiferromagnetic interaction can mitigate the detrimental effects of the bath. Additionally, as $\alpha$ increases, we observe that the fidelity remains very high, as supported in the Appendix~\ref{app:bathcoup}, where we show fidelity and leakage for various $\alpha$ couplings.\\
One of the limitations of our work is the inevitable ``loss" of information that occurs when encoding one qubit in two physical ones, as it requires the use of two qubits to store the information of a single qubit. Another limitation lies in determining how to model the interaction with the bath, a common critique for encoding strategies proposed in the existing literature. However, the potential ``gain" achieved through a non-detrimental bath effect on the system dynamics could counterbalance the aforementioned ``loss" of information. It is worth noting that the long-wavelength limit \cite{fradkin2013field} holds when dealing with two qubits, justifying the use of the same coupling method, global bath, and coupling constant. Yet, this assumption may become problematic when dealing with a chain of more than two qubits.\\
Since the encoding can be done a posteriori, utilizing the mean values of the observables of the two physical qubits, the experimental implementation only requires an antiferromagnetic interaction between the qubits. One possible approach for this implementation is the use of a coupler \cite{blais2021circuit, yan2018tunable} (qubit or electromagnetic mode) that achieves an effective interaction between the two qubits without directly coupling them. \\
Our work helps to understand and analyze the potential of the interaction between qubits in the presence of the bath. As a future study, we could investigate the behavior when more qubits are involved, since the DFSs become more intriguing. Additionally, we can conduct a detailed analysis of the model by introducing a coupler through MPS numerical simulations, accurately reproducing the system dynamics.

\begin{acknowledgments}
G.D.F. and A.H. acknowledge financial support from 376 PNRR MUR Project No. PE0000023-NQSTI; A.H. was also supported by the PNRR MUR Project CN $00000013$-ICSC. C.A.P. acknowledges founding from the IQARO (SpIn-orbitronic QuAntum bits in Reconfigurable 2D-Oxides) project of the European Union's Horizon Europe research and innovation programme under grant agreement n. 101115190. C.A.P. acknowledges founding from the PRIN 2022 PNRR project P2022SB73K - Superconductivity in KTaO3 Oxide-2DEG NAnodevices for Topological quantum Applications (SONATA). G.D.F. and C.A.P. acknowledge financial support from the PRIN 2022 project 2022 FLSPAJ “Taming Noisy Quantum Dynamics” (TANQU). The authors acknowledge interesting discussions with A. de Candia.
\end{acknowledgments}

\appendix
\section{Lindblad vs MPS and decay rates estimation}\label{app:lindblad}
\begin{figure*}[t]
    \begin{center}
        \includegraphics[scale=0.325]{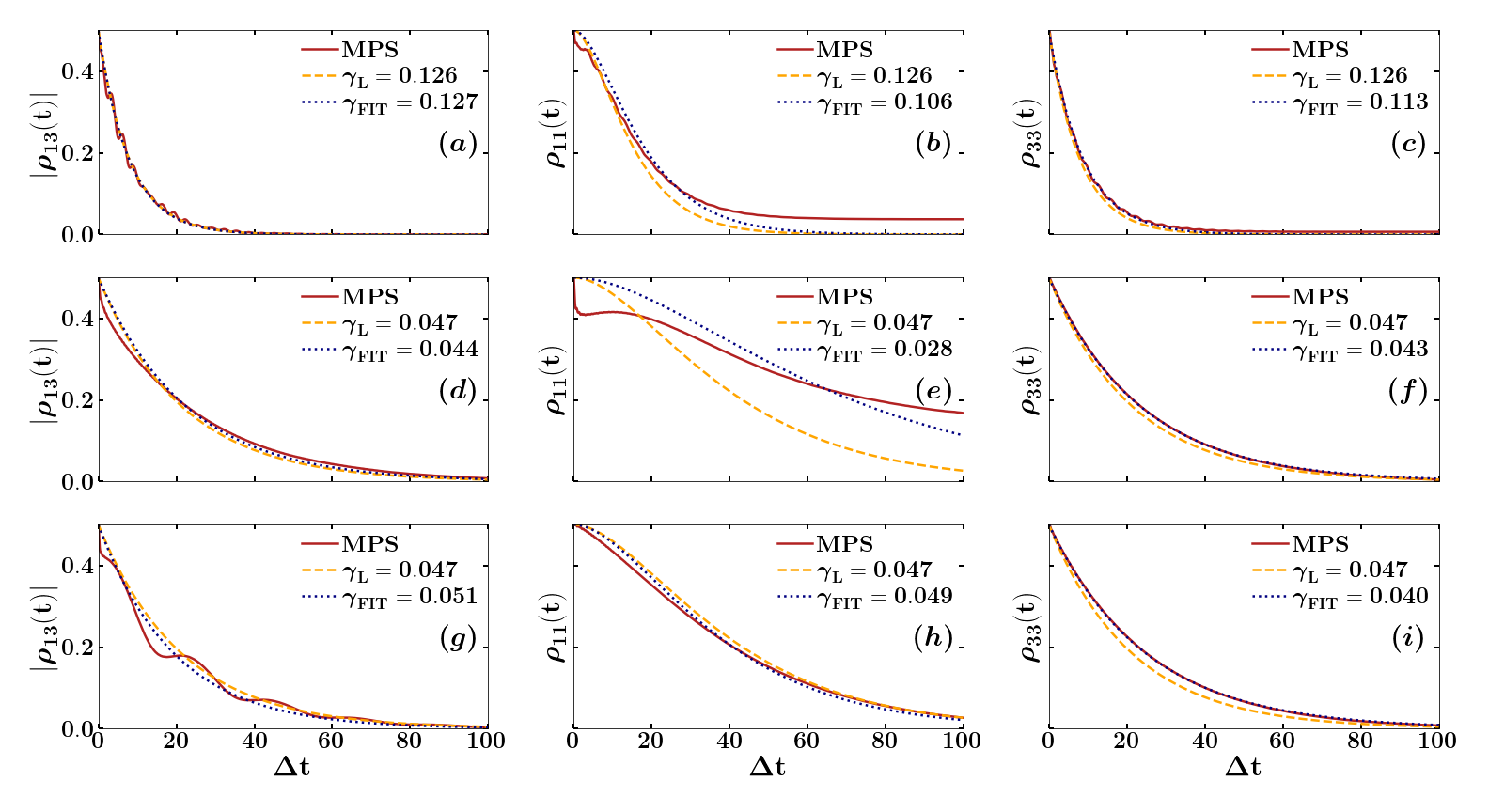} 
        \caption{\label{fig:confr_lind}Coherence modulus $|\rho_{13}(t)|$ and populations $\rho_{11}(t)=P_1$, and $\rho_{33}(t)=P_3$ of the energy eigenstates $\ket{1}$, and $\ket{3}$ of $H_0$. These are the only non-zero terms of the reduced density matrix $\rho(0)$, given the initial state (Eq.~\ref{eq:initstate}), displayed as functions of dimensionless time $\Delta t$. Also, the population $\rho_{00}(t)=P_0$ is fixed by normalization. The bath coupling strength is $\alpha=0.02$, and the qubit-qubit interaction strength takes the values $\nu=0\Delta$ in the first line, $\nu=5\Delta$ in the second line, and $\nu=-5\Delta$ in the third line. In the plot, the red solid lines correspond to the MPS numerical results, while the orange dashed lines are the Lindblad master equation solutions and the blue dotted lines are the fits obtained by using the same functional forms as the Lindblad solutions.} 
    \end{center}
\end{figure*}
In this appendix we wish to clarify that MPS numerical simulations are necessary instead of relying on the more commonly used Lindblad master equation solution. We focus on the relaxation and decoherence due to the presence of the bath \cite{gargiulo2011electronic,perroni2011spectral,perroni2013single,perroni2014noise,perroni2015interplay}. It's worth noting that the singlet state $\ket{\psi_S}$ is DFS due to the nature of its interaction with the bath, which means that it is not affected by decoherence.\\
We follow the notation used in \cite{schaller2014open} for the Lindblad master equation in presence of the interaction Hamiltonian $H_{qub-bath}=A\otimes B$, where $A=\sigma_z^1+\sigma_z^2$ is the system operator coupled to the bath one $B=\sum_{i=1}^N\lambda_i\left(a_i+a_i^{\dagger}\right)$. The singlet state does not participate in the dynamics and can be effectively decoupled from the other equations. Hence, in the interaction picture using the system energy eigenbasis $H_{qub}\ket{i}=E_i\ket{i}$, with $i=0,\dots,3$ and neglecting the Lamb-shift Hamiltonian, the Lindblad equation reads:
\begin{align}
    \frac{d\rho}{dt}&=\gamma_{01,01}\left[\ket{0}\bra{1}\rho(t)\ket{1}\bra{0}-\frac{1}{2}\left\{\ket{1}\bra{1},\rho(t)\right\}\right]\nonumber\\
    &\,+\gamma_{13,13}\left[\ket{1}\bra{3}\rho(t)\ket{3}\bra{1}-\frac{1}{2}\left\{\ket{3}\bra{3},\rho(t)\right\}\right],
\end{align}
where the decay rates are:
\begin{align}
    \gamma_{01,01}&=\gamma(E_1-E_0)\bra{0}(\sigma_z^1+\sigma_z^2)\ket{1}\bra{1}(\sigma_z^1+\sigma_z^2)\ket{0}\nonumber\\
    &=4\left(b\left(\frac{\nu}{\Delta}\right)\right)^2 2\pi J(E_1-E_0),\\
    \gamma_{13,13}&=\gamma(E_3-E_1)\bra{1}(\sigma_z^1+\sigma_z^2)\ket{3}\bra{3}(\sigma_z^1+\sigma_z^2)\ket{1}\nonumber\\
    &=4\left(a\left(\frac{\nu}{\Delta}\right)\right)^2 2\pi J(E_3-E_1).
\end{align}
Here $J(E_b-E_a)$ is the spectral density of the bath at the energy difference $E_b-E_a$, and $a(\nu/\Delta)$ (Eq.~\ref{eq:afac}) and $b(\nu/\Delta)$ (Eq.~\ref{eq:bfac}) are the coefficients in the definitions of the system eigenstates. We can use these equations to calculate the time-evolution of the populations and coherences of the system reduced density matrix. If we look at the populations $\rho_{00}(t)$ and $\rho_{11}(t)$ we see that there is the difference between the two decay rates. Furthermore, if we use an Ohmic spectral density $J(\omega)$ (Eq.~\ref{eq:specdens}) with a cutoff made by the Heaviside function, we obtain a $0/0$ indeterminate form. Hence, we choose to model the system decay using an exponential function with the cutoff frequency and then take the limit as the cutoff frequency goes to infinity. We only need to compute one limit, for example, for $\rho_{11}(t)$ and $\rho_{00}(t)$ will be fixed by the normalization ($\sum_{i=0}^3 \rho_{ii}(t)=1$). By defining a single decay rate $\gamma=2\pi\alpha c$, where the constant $c=(a(\nu/\Delta))^2(\sqrt{4\Delta^2+\nu^2}+\nu)=(b(\nu/\Delta))^2(\sqrt{4\Delta^2+\nu^2}-\nu)$, the populations and coherences can be rewritten in a very compact way as follows:
\begin{align}
    \rho_{00}(t)&=1-\rho_{11}(0)e^{-\gamma t}-\rho_{22}(0)-\rho_{33}(0)e^{-\gamma t}(1+\gamma t)\label{eq:rho00}\\
    \rho_{11}(t)&=\rho_{11}(0)e^{-\gamma t}+\rho_{33}(0)e^{-\gamma t}\gamma t \\
    \rho_{22}(t)&=\rho_{22}(0)\\
    \rho_{33}(t)&=\rho_{33}(0)e^{-\gamma t}\\
    \rho_{01}(t)&=\rho_{01}(0)e^{-\frac{\gamma}{2}t}\\
    \rho_{03}(t)&=\rho_{03}(0)e^{-\frac{\gamma}{2}t}\\
    \rho_{13}(t)&=\rho_{13}(0)e^{-\gamma t}\label{eq:rho13}.    
\end{align}
It's worth noting that the coherence $\rho_{13}$ exhibits a decay rate equal to $\gamma$, which is the same as that of the population $\rho_{33}$. Meanwhile, the decay rate of the other two coherences is half of $\gamma$, similar to what occurs in the case of a single qubit. Now, we can use these time functions to compare them with our MPS numerical results and estimate the accuracy of the Lindblad solution in predicting the decay rates, as well as the errors that may arise in the dynamics when using it, starting from the quasi-ferromagnetic (F) state:
\begin{equation}
    \label{eq:initstate}\ket{\psi(0)^{F}}=\frac{\ket{1}+e^{i\pi/4}\ket{3}}{\sqrt{2}}\otimes\ket{0,\dots,0}_B.
\end{equation}
This means that the two-qubit system is initially in a combination of the most excited ferromagnetic eigenstates of the closed Hamiltonian ($\ket{1}$ and $\ket{3}$), and the bath is empty at zero temperature. \\
We investigate three values of the interaction strength between the qubits, $\nu=\{-5,0,5\}\Delta$, to study the interplay between the energy scale of the system and the bath effects. Additionally, we observe the dynamical behavior of the system for two different values of the coupling to the bath, $\alpha=\{0.01,0.02\}$.\\
Figure~\ref{fig:confr_lind} displays the time evolution of the modulus of the coherence $|\rho_{13}(t)|$ and the populations $\rho_{11}(t)=P_1$, and $\rho_{33}(t)=P_3$. These populations are computed for a typical bath coupling strength of $\alpha=0.02$, which is of the same order of magnitude as the coupling strength used in the main text. The density matrix elements are shown for the three different values of the qubit-qubit interaction strength: $\nu=0\Delta$ in the first line, $\nu=5\Delta$ in the second line, and $\nu=-5\Delta$ in the third line.\\
It is evident that the Lindblad equation, which is a perturbative approach in the coupling to the bath and does not consider higher-order processes, only accurately predicts the exponential decay of the $P_3$ population. Nevertheless, it fails to correctly forecast the population $P_1$ and $P_0$ (not displayed in the figure, but fixed by normalization) as higher-order processes become significant and are not taken into account.\\
For the coherence modulus, we observe that the Lindblad solution cannot reproduce the oscillations of the MPS numerical simulations. Additionally, for $\nu\neq0\Delta$, it also predicts an incorrect initial transient. Looking at the real and imaginary parts of the coherence separately (not shown in the figure for clarity), we observe more discrepancies in frequencies and amplitudes of the oscillations, which are somewhat balanced when considering the modulus. To estimate the decay rates of the MPS results $\gamma_{FIT}$, we also plot the results obtained by fitting the MPS data with the same functional forms as the Lindblad equation solutions (Eqs.~from \ref{eq:rho00} to \ref{eq:rho13}).\\
It's worth noting that the ferromagnetic qubit-qubit interaction (the second line in Figure~\ref{fig:confr_lind}), especially for $P_1$, enhances the effect of the bath, which always induces the ferromagnetic order of the qubits. As a result, the Lindblad equation fails to reproduce the expected stationary state. On the other hand, the antiferromagnetic qubit-qubit interaction (the third line in Figure~\ref{fig:confr_lind}) offers some protection against the bath and yields the correct stationary state, even for this intermediate value of $\alpha$. This suggests that the particular type of qubit-qubit interaction can significantly impact the behavior of the system in the presence of a bath.

\section{Fidelity in the logical subspace of best encoding strategies}\label{app:fidaf}
\begin{figure}[H]
    \begin{center}
        \includegraphics[scale=0.26]{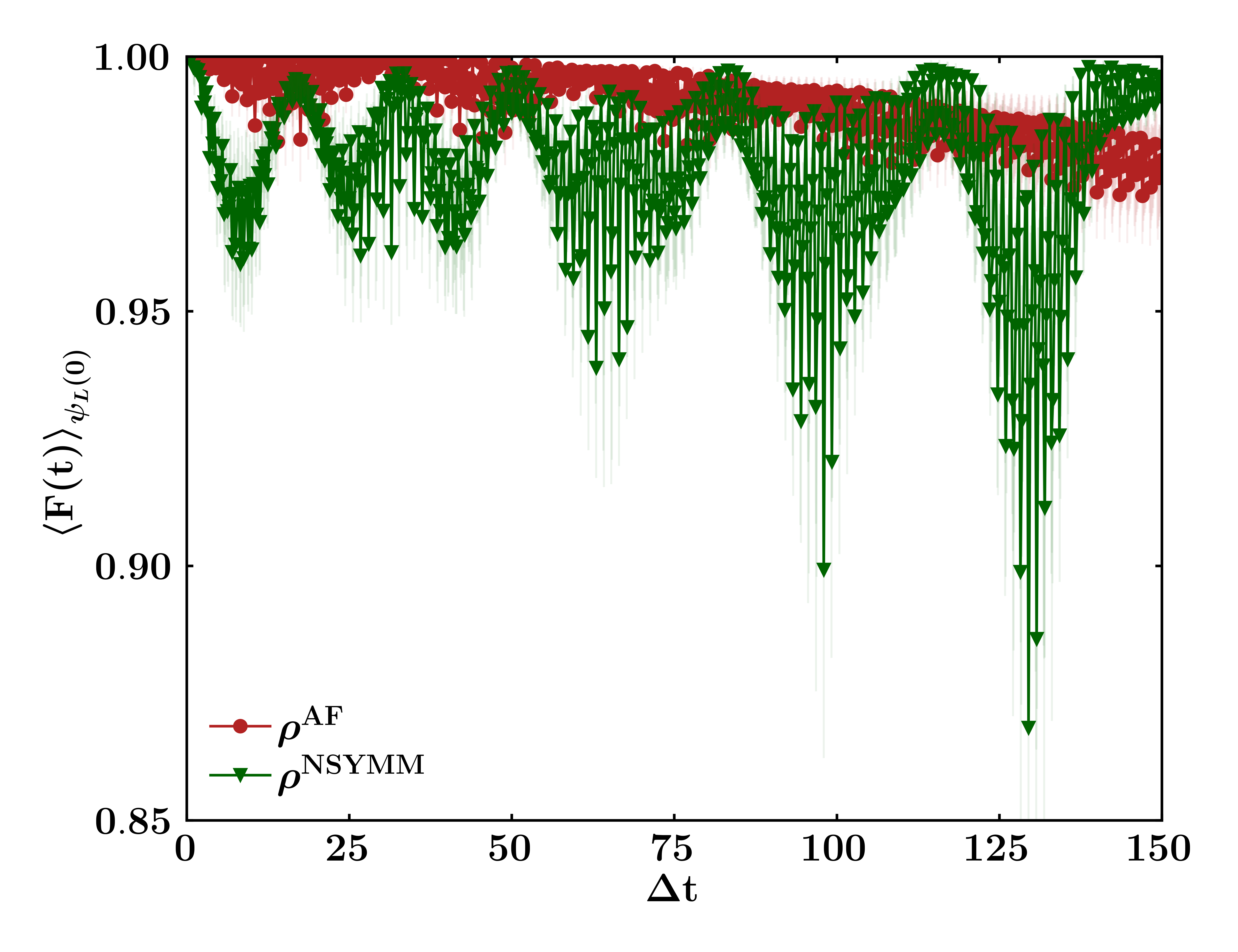}  
        \caption{\label{fig:fid_enc}Fidelity $F(t)$ of the free evolution with the open system evolution of encoded qubits AF (red circle), and NSYMM (green triangle) as a function of dimensionless time $\Delta t$ for $\nu=-5\Delta$, $\alpha=0.01$, and the initial state $\ket{\psi_L(0)}=\cos{\theta}\ket{\uparrow_L}+e^{i\phi}\sin{\theta}\ket{\downarrow_L}$ where $\theta$ and $\phi$ sample all the logical qubit Hilbert space. We average the fidelity over $18$ realizations of these angles. The dots in the plot correspond to the average value of the fidelity, while the shaded regions around them shows the range of values covered by the standard deviation of the fidelity.} 
    \end{center}
\end{figure}

In this appendix, we aim to emphasize that the AF strategy remains the superior choice when considering the new logical subspace of the qubit. This observation is underscored by the data presented in Figure~\ref{fig:fid_enc}. Specifically, the NSYMM strategy exhibits a fidelity that oscillates with increasing amplitudes over time, reaching a value of approximately 0.85 within the duration of our simulation. Conversely, the AF strategy displays smaller oscillation amplitudes and approaches a fidelity value of around 0.97. \\
This further supports the conclusion that the AF strategy is the preferred option for maintaining qubit integrity within the new logical subspace.

\section{Effects on the encoding of increasing the system-bath coupling strength} \label{app:bathcoup}
In this appendix, we investigate how increasing the coupling strength of the system to the bath affects the optimal encoding strategy. Specifically, we explore how the variation of $\alpha$ impacts the fidelity and leakage of the encoded information, and determine whether the encoding strategy becomes ineffective under stronger coupling regimes. Our results provide valuable information on the robustness of the encoding scheme under different coupling conditions.\\
\begin{figure}[H]
    \begin{center}
        \includegraphics[scale=0.26]{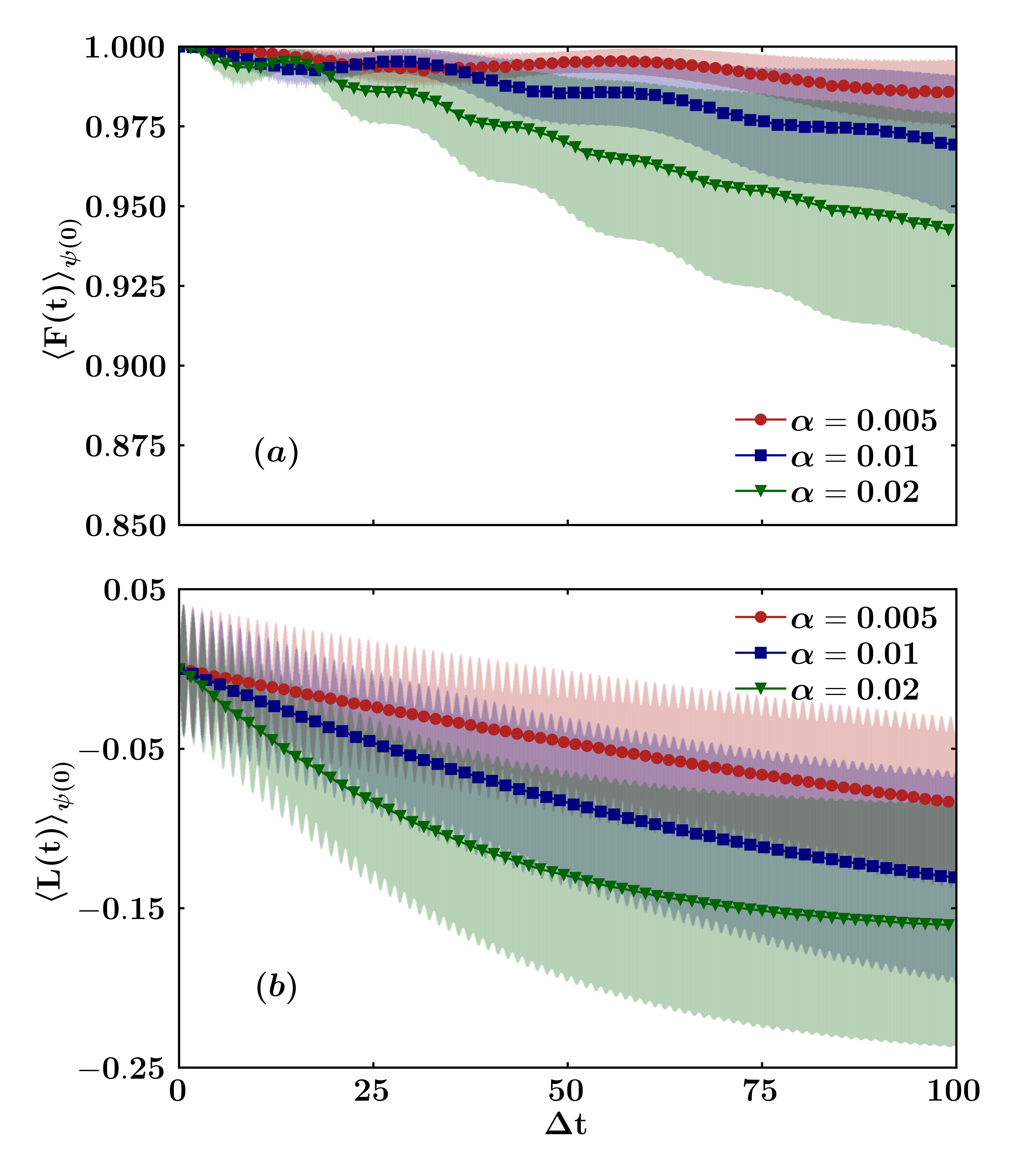}  
        \caption{\label{fig:fid_leak_bestenc}Fidelity $F(t)$ of the free evolution with the open system evolution ($a$) and leakage $L(t)$ ($b$) of the encoded qubit AF as functions of dimensionless time $\Delta t$ for $\nu=-5\Delta$, $\alpha=0.005$ (red circle), $\alpha=0.01$ (blue square), $\alpha=0.02$ (green triangle) and the initial state $\ket{\psi(0)}=d_S\ket{S}+d_{T,AF}\ket{T,AF}+d_{T,F+}\ket{T,F+}+d_{T,F-}\ket{T,F-}$, where $d_S$, $d_{T,AF}$, $d_{T,F+}$ and $d_{T,F-}$ sample all the two-qubits Hilbert space. We average the fidelity and the leakage over $332$ realizations of these coefficients. The dots in the plots correspond to the average value of the fidelity and the leakage, while the shaded regions around them show the range of values covered by the standard deviation of the fidelity and the leakage.} 
    \end{center}
\end{figure}
Figure~\ref{fig:fid_leak_bestenc} provides insight into the robustness of the encoding paradigm concerning the coupling strength parameter $\alpha$. In Figure~\ref{fig:fid_leak_bestenc}.a, we observe the efficacy of our chosen encoding strategy, which consistently approaches high asymptotic values, even as $\alpha$ increases. Furthermore, Figure~\ref{fig:fid_leak_bestenc}.b illustrates that as $\alpha$ increases, the degree of leakage becomes more negative. This behavior is attributed to the increased influence of the bath on the system, particularly as the system approaches its ground state. Thus, our chosen encoding strategy exhibits strong resilience to the detrimental effects of the bath, in contrast to that of physical qubits. Moreover, the fidelity values are even higher, indicating that we are effectively leveraging the interaction between the two qubits to counteract the bath's negative effects.
\section{Lindblad fidelity dynamics of the Bell states}
\begin{figure*}[t]
\centering
    \includegraphics[scale=0.3]{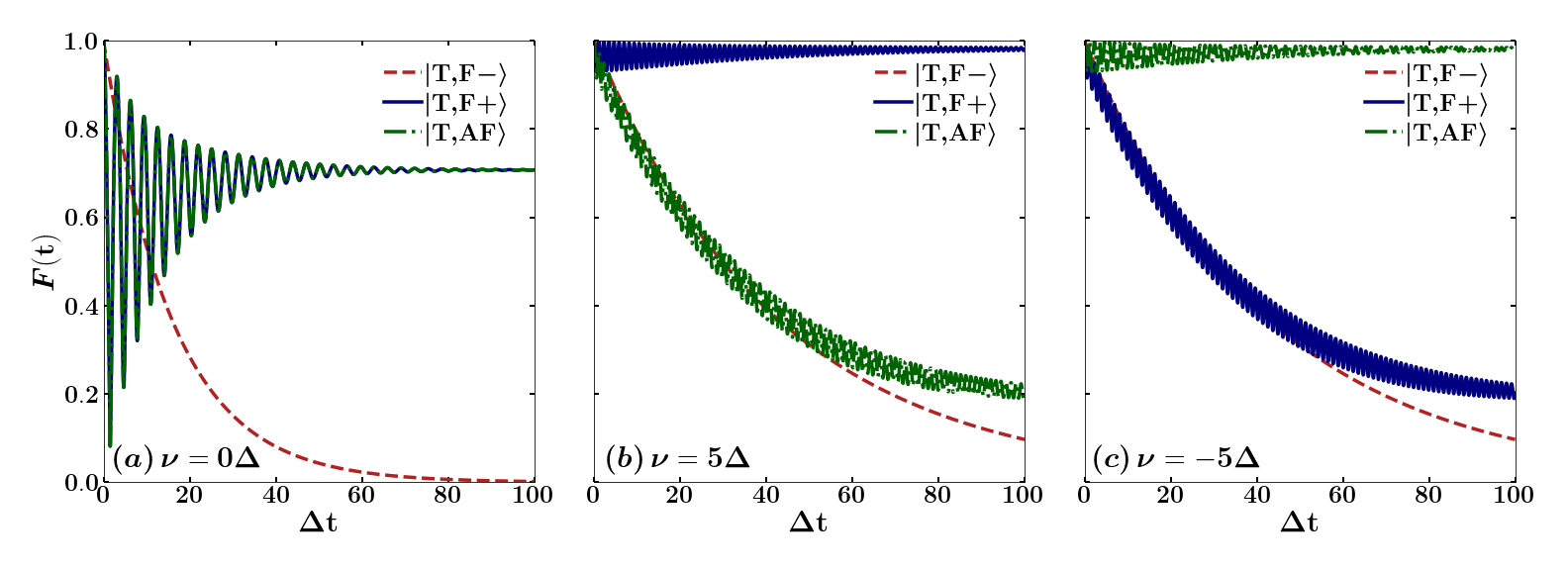}
    \caption{\label{fig:fid_lind}Fidelity $F(t)$ of the free ($\alpha=0$) evolution of the three initial states in the triplet ${\ket{T,F-},\ket{T,F+},\ket{T,AF}}$ (respectively red dashed, solid blue and solid-dashed green lines) with the open system evolution as a function of dimensionless time $\Delta t$ for $\alpha = 0.02$ and $\nu = 0\Delta$ ($a$), $\nu = 5\Delta$ ($b$) and $\nu = -5\Delta$ ($c$), computed with the analytical Lindblad approach.}
\end{figure*}

In the main text, we assert that the antiferromagnetic state is particularly resilient due to the antiferromagnetic interaction $\nu$ (see Eq.~\ref{eq:Hqub}), hence forming the core of our antiferromagnetic strategy. The fundamental concept underlying this assertion is that when the interaction is antiferromagnetic at very high $|\nu|>>\Delta$, the antiferromagnetic state closely coincides with the ground state, suggesting its high resilience against interactions with the environment. Here, we aim to validate this concept by computing an analytical expression for fidelity dynamics. To achieve this, we calculate the time-evolved density matrices, both without and with the bath, using the Lindblad formalism, initializing the system in one of three states within the triplet ${\ket{T,F-},\ket{T,F+},\ket{T,AF}}$. Subsequently, we evaluate the fidelity for these three cases. The calculations are simplified since the state of the closed system is pure. Hence, the fidelity reduces to $F\left[\rho^o(t),\rho^c(t)\right]=\sqrt{\bra{\psi^c(t)}\rho^o(t)\ket{\psi^c(t)}}$. Here, as in the main text, $\rho^o$ is the density matrix of the open system and $\rho^c(t)$ that of the closed one (without bath interaction). The analytical expressions for the fidelity of the three initial states in the triplet are as follows:
\onecolumngrid
\begin{align*}
    &F\left[\rho^o_{T,F-}(t),\rho^c_{T,F-}(t)\right]=e^{-\frac{\gamma}{2}t}\\
    &F\left[\rho^o_{T,F+}(t),\rho^c_{T,F+}(t)\right]=\sqrt{b^2+(a^4-a^2 b^2 (1+\gamma t))e^{-\gamma t}+2 a^2 b^2 e^{-\frac{\gamma}{2}t} \cos[(E_0-E_3)t]}\\
    &F\left[\rho^o_{T,AF}(t),\rho^c_{T,AF}(t)\right]=\sqrt{a^2+(b^4-a^2 b^2 (1+\gamma t))e^{-\gamma t}+2 a^2 b^2 e^{-\frac{\gamma}{2}t}\cos[(E_0-E_3)t]},
\end{align*}
\twocolumngrid
where $a$ and $b$ are defined in Eqs.~\ref{eq:afac} and \ref{eq:bfac}, $E_i$ are the eigenvalues of the two-qubit Hamiltonian in Eq.~\ref{eq:eigen} and the decay rate $\gamma $ is the one obtained in Appendix \ref{app:lindblad}. It's worth noticing that Lindblad approach allows us to write down an analytical compact form for the fidelity of the stationary state of the antiferromagnetic state, which is $F\left[\rho^o_{T,AF}(t),\rho^c_{T,AF}(t)\right]=|a|$. Additionally, under the condition of an antiferromagnetic interaction where $|\nu|>>\Delta$, the fidelity of $\ket{T,F+}$ mirrors the time evolution of $\ket{T,F-}$, while the antiferromagnetic state remains consistently at a fidelity value of $1$, reinforcing our hypothesis. In Figure~\ref{fig:fid_lind}, we illustrate the fidelities of these three states for the three interaction values $\nu\in[-5,0,5]\Delta$. Across all scenarios, the fidelity of the state $\ket{T,F-}$ (independent of $\nu$) decays, while the other two oscillate before stabilizing at a fixed value. Specifically, in the absence of interaction, the behaviors of the other two states align, whereas in the remaining scenarios, their behaviors interchange. For the ferromagnetic case, the most resilient state is the ferromagnetic one, while for the antiferromagnetic case, it is the antiferromagnetic state. It's worth noting that this symmetric behavior holds true only within the Lindblad approach; the MPS method captures the differences between these behaviors.

\section{MPS numerical simulations}\label{app:MPS}
We employed time-dependent matrix product state (MPS) simulations, implemented with ITensor Library \cite{fishman2022itensor}, to investigate the system's dynamics. We analyzed the fidelity, the leakage and the purity of the encoded system over time. The long-range interactions between each of the two qubits and the each bath mode were described using the star geometry. In this configuration, the qubits of frequency $\Delta$ were placed on the first two sites, and the collection of $N$ bosonic modes of the bath with frequencies $\omega_i$ on the remaining sites. The couplings $\lambda_i$ between each qubit and each bosonic mode were defined to describe the bath in terms of an Ohmic spectral density, as explained in the main text. The coupling constants to the bath are $\left|\lambda_i\right|=\sqrt{\frac{q_0 k_i\omega_i}{8}}=\frac{\omega_c}{N}\sqrt{\alpha i}$, where $\omega_c$ is the cutoff frequency, $k_i=m_i\omega_i^2$ and $q_0$ is the position of the minima in the symmetric double well potential from which we derive the spin-boson Hamiltonian for the qubits, that we set to $1$. We neglected the energy shift $\sum_{i=1}^N \omega_i/2$, which does not affect the dynamics. The dimensionless parameter $\alpha$ measures the strength of the qubits-bath coupling.\\
We studied the system's dynamics for different values of $\alpha$ in the range $[0.005, 0.2]$, of their interaction $\nu=[-5,0,5]\Delta$ and setting $\omega_c = 10\Delta$. We selected $332$ realizations of the initial state for simulating the system's dynamics. We applied the time-dependent variational principle (TDVP) \cite{haegeman2011time,haegeman2016unifying,paeckel2019time}, where the time-dependent Schrödinger equation is projected onto the tangent space of the MPS manifold with a fixed bond dimension at the current time.\\
In this study, we employed the two-site TDVP (2TDVP as described in \cite{paeckel2019time}), using a second-order integrator with a left-right-left sweeping approach and a half-time step of $dt/2$. This method exhibits a time-step error of $O(dt^3)$, with accuracy controlled by the MPS bond dimension and the threshold to terminate the Krylov series. We halted the Krylov vectors recurrence when the total contribution of two consecutive vectors to the matrix exponential dropped below $10^{-12}$. While more advanced methods, such as basis extension optimization \cite{zhang1998density,brockt2015matrix}, exist, we opted for convergence in the number of bath modes ($N = 250$) with Hilbert space dimension ($N_{bos} = 3$). This approach allowed us to find the optimal compromise between the smallest bond dimension and longest simulation times, by converging also over the time interval that we set to $\delta dt=0.05$.\\
Our truncation error remained below $10^{-13}$ by requiring a maximum bond dimension of $D_{max} = 50$. Simultaneously, this optimal maximum bond dimension enabled us to achieve a final time for our simulations as large as $\Delta t_{final}=150$.

\bibliography{biblio}% Produces the bibliography via BibTeX.

\end{document}